\PassOptionsToPackage{usenames,dvipsnames}{xcolor}

\documentclass[manuscript,screen,acmsmall]{acmart} 
\usepackage{etoolbox}
\usepackage{soul} 
\usepackage{enumitem}

\usepackage{pifont} 
\usepackage{multirow}
\usepackage{array} 
\usepackage{longtable}
\usepackage{booktabs} 
\newcolumntype{P}[1]{>{\centering\arraybackslash}p{#1}}
\newcolumntype{M}[1]{>{\centering\arraybackslash}m{#1}}

\usepackage{rotating}
\usepackage{tabularx}

\usepackage{tikz}
\newcommand{\rqtwo}[1]{\tikz[baseline={(a.base)}]\node[draw=green!75!black, line width=0.7pt, rounded corners=0.8ex, fill=green!10!white, inner sep=1.5pt,text=black](a){#1};}

\newcommand{\rqthree}[1]{\tikz[baseline={(a.base)}]\node[draw=orange!75!black, line width=0.7pt, rounded corners=0.8ex, fill=orange!10!white, inner sep=2pt, text=black](a){#1};}

\newcommand{\rqfour}[1]{\tikz[baseline={(a.base)}]\node[draw=pink, line width=0.7pt, rounded corners=0.8ex, fill=pink!25!white, inner sep=2pt, text=black](a){#1};}

\newcommand{\rqfive}[1]{\tikz[baseline={(a.base)}]\node[draw=blue!40!white, line width=0.7pt, rounded corners=0.8ex, fill=blue!10!white, inner sep=2pt, text=black](a){#1};}

\usepackage{hyperref}
\usepackage{multirow}

\usepackage{caption}

\usepackage{pifont}
\usepackage[usenames,dvipsnames]{xcolor}
\newcommand{\greencheck}{{\color{ForestGreen}\checkmark}}
\newcommand{\xmark}{\color{red}\ding{55}}%

\AtBeginDocument{%
  }

\setcopyright{acmlicensed}
\acmJournal{CSUR}
\copyrightyear{2024}
\acmYear{2024}
\acmDOI{XXXXXXX.XXXXXXX}

\begin{document}

\title{Frameworks, Modeling and Simulations of Misinformation and Disinformation: A Systematic Literature Review}


\author{Alejandro~Buitrago~L\'opez}
\email{alejandro.buitragol@um.es}
\orcid{0009-0002-1606-8766}
\affiliation{%
  \institution{Department of Information and Communications Engineering, University of Murcia}
  \city{Murcia}
  \country{Spain}
}

\author{Javier~Pastor-Galindo}
\email{javierpg@um.es}
\orcid{0000-0003-4827-6682}
\affiliation{%
  \institution{Department of Information and Communications Engineering, University of Murcia}
  \city{Murcia}
  \country{Spain}
}

\author{Jos\'e~A.~Ruip\'erez-Valiente}
\email{jruiperez@um.es}
\orcid{0000-0002-2304-6365}
\affiliation{%
  \institution{Department of Information and Communications Engineering, University of Murcia}
  \city{Murcia}
  \country{Spain}
}

\renewcommand{\shortauthors}{Buitrago~L\'opez, et al.}

\begin{abstract}
The prevalence of misinformation and disinformation poses a significant challenge in today's digital landscape. That is why several methods and tools are proposed to analyze and understand these phenomena from a scientific perspective. To assess how the mis/disinformation is being conceptualized and evaluated in the literature, this paper surveys the existing frameworks, models and simulations of mis/disinformation dynamics by performing a systematic literature review up to 2023. After applying the PRISMA methodology, 57 research papers are inspected to determine (1) the terminology and definitions of mis/disinformation, (2) the methods used to represent mis/disinformation, (3) the primary purpose beyond modeling and simulating mis/disinformation, (4) the context where the mis/disinformation is studied, and (5) the validation of the proposed methods for understanding mis/disinformation. 

The main findings reveal a consistent essence definition of misinformation and disinformation across studies, with intent as the key distinguishing factor. Research predominantly uses social frameworks, epidemiological models, and belief updating simulations. These studies aim to estimate the effectiveness of mis/disinformation, primarily in health and politics. The preferred validation strategy is to compare methods with real-world data and statistics. Finally, this paper identifies current trends and open challenges in the mis/disinformation research field, providing recommendations for future work agenda.
\end{abstract}




\maketitle

\section{Introduction}
Individuals navigate in a world inundated with an unprecedented volume of information. The rise of the internet has fundamentally changed how citizens become informed, how they discuss and how they form their opinions \cite{Quattrociocchi2014}. Online Social Networks (OSNs) play a pivotal role in this new information ecosystem, allowing individuals to disseminate information globally easily. However, the democratization of information comes with inherent challenges, as the content shared on these platforms may not always be accurate, giving rise to misinformation and disinformation \cite{doi:10.1126/science.aao2998}. Particularly, the emergence of an interconnected digital environment, where data moves without restraint among various platforms, has brought to light a pressing concern: the exploitation of information to deceive, mislead or shape public opinion.

Misinformation and disinformation (mis/disinformation) existed before OSNs, but the diffusion of false information is a phenomenon that has gained significant notoriety in the digital era, primarily due to the facilities for creating and disseminating content offered by OSNs, being a major threat to democracy, information-seeking, open debate, and a free and modern society \cite{enisa2022}. The propagation of false information in OSNs affects public perception of important events\cite{9226407}, undermining trust in institutions and affecting political decision-making \cite{van2017beating}. Another example may be that of fake news in the healthcare setting. This can have serious consequences, leading people to make decisions that put their health or even their lives at risk, may delay appropriate treatment, and worsen the prognosis of serious diseases \cite{poland2010fear}.

The escalation in the popularity of disinformation materialized in 2016, when the term ``post-truth'' was coined as the ``word of the year'' by Oxford Languages, reflecting when emotions and personal beliefs replace objective facts \cite{oupOxfordWord}, emphasizing the multidisciplinary nature of the phenomenon. From a social aspect and analyzing people's behavior in OSNs, previous works have found a cognitive phenomenon called confirmation bias, in which people tend to interpret, remember or favor information that confirms their pre-existing beliefs or hypotheses \cite{bessi2015trend}. Since individuals more inclined to believe such information are similarly more likely to share it, this fact is related to another previous finding which analyzed that negative messages spread faster than positive ones \cite{ferrara2015quantifying}. This phenomenon results in echo chambers emerging in OSNs, where individuals are exposed to information that aligns with their beliefs, reinforcing preconceptions and limiting exposure to alternative perspectives \cite{echochambers}. This context has been recently exploited through a great wave of disinformation attacks that anticipated and accompanied events such as the Russia-Ukraine war \cite{doi:10.1080/17512786.2016.1163237}, the Israel-Palestine conflict \cite{apnewsMisinformationAbout} or the COVID-19 pandemic \cite{cinelli2020covid}.

Gaining attention from the research community, governments and the public at large, the study and modeling of mis/disinformation leads to the utilization of frameworks, models and simulations to comprehend and conceptualize this phenomenon. Frameworks provide high-level structured and organized schemes to structure processes, concepts or dimensions of mis/disinformation \cite{Agarwal2021}. Among the different approaches, technical ones are usually oriented to support the detection and combat disinformation \cite{Terp2022}, recently including AI components~\cite{feuerriegel2023research}. Models are formal proposals that explain processes or activities of mis/disinformation dynamics, such as utilizing epidemic models \cite{10.4018/IJSWIS.300827}. Finally, simulations are computational executions that understand and evaluate the operation of certain mis/disinformation processes, usually designed under certain frameworks or models. Thanks to advances in artificial intelligence (AI), they do not necessarily have to be based on real data or people \cite{s3}.

Despite the existing reviews in the field of misinformation and disinformation, we have not found any specific frameworks, modeling or simulations. For example, the authors in \cite{slr1} and in \cite{Capuano2023} review fake news detection models contributed by various machine learning and deep learning algorithms categorized and described in different datasets. Moreover, in \cite{Zhou2020}, a comprehensive analysis is provided of the various techniques employed for rumor and fake news detection and not only considers AI techniques. On the other hand, Kapantai \textit{et al.} \cite{Kapantai2021} provided a systematic review of 10 experimental studies to create a unified taxonomic framework for understanding and categorizing different types of disinformation. Finally, Gupta \textit{et al.} \cite{Gupta2022} reviewed 92 research articles to analyze existing solutions in the fight against disinformation and propose a potential solution to combat fake news through a news verification service. 

The surveys mentioned above focus on detection methodologies, particularly using AI, and move around the concept of fake news, a very limited vision of disinformation. Given the rapid proliferation of misleading information in various domains, understanding and researching these phenomena has gained significance. In this sense, our goal is to unveil how mis/disinformation is conceptualized, modeled and simulated in the literature in studies that aim to understand, research and assess mis/disinformation dynamics and processes. This situation calls for a comprehensive review of the current state of the art on frameworks, models and simulations around misinformation/disinformation. The current paper aims to conduct the first systematic literature review in these respects and answer the following research questions:

\begin{enumerate}[label=\textbf{RQ\arabic*.}]
  \item What are key commonalities and differences in defining misinformation and disinformation?
  \item How has the mis/disinformation phenomenon been analyzed, modeled and simulated in the literature?
  \item What purposes motivate the existing frameworks, models and simulations around mis/disinformation?
  \item Which contexts are addressed by the existing frameworks, models and simulations around mis/disinformation?
  \item Which validation is performed in the existing frameworks, models and simulations around mis/disinformation?
\end{enumerate}

The rest of the paper is organized as follows. Section \ref{methodology} describes the methodology followed in the systematic review, including some terminology clarifications, the research questions, databases and search terms, research selection and review process. Section \ref{results} presents the analysis and synthesis of the survey results. Section \ref{discussion} includes an analysis of our findings by extracting the current trends and open challenges. Finally, Section \ref{conclusion} incorporates the conclusions of this work.

\section{Methodology}\label{methodology}

\begin{figure}[ht!]
    \centering
    \includegraphics[width=0.9\textwidth]{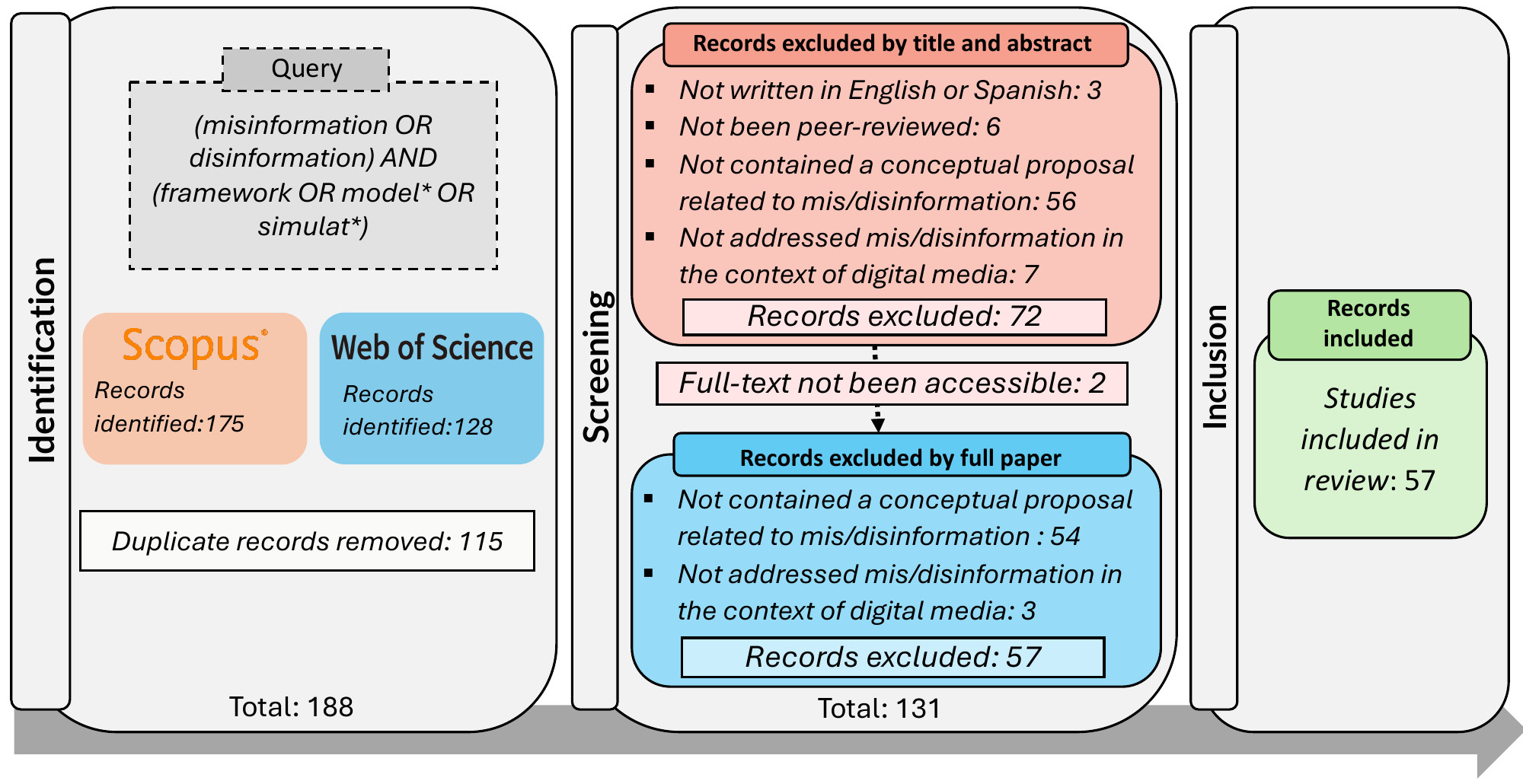}
    \caption{PRISMA flow diagram with the following methodology.}
    \label{fig:prisma}
\end{figure}

A standard systematic literature review methodology was followed, using the \textit{Preferred Reporting Items for Systematic Reviews and Meta-Analyses} (PRISMA) \cite{Pagen160} as a basis for conducting our study. Figure \ref{fig:prisma} shows the PRISMA diagram representing the different stages of our systematic review (inspired by the original proposal \cite{Pagen71}) and is explained below. 

\begin{itemize}
    \item Definition of key concepts and formulation of each RQ.
    \item Application of search queries to pre-identified bibliographical databases.
    \item Definition and application of a set of inclusion and exclusion criteria.
    \item Conducted a comprehensive paper review and coding process for the research questions, followed by synthesis and analysis.
\end{itemize}

\subsection{Definition of key concepts and research questions} \label{rqs}

\begin{figure*}[ht]
\begin{center}
\includegraphics[width=0.8\linewidth]{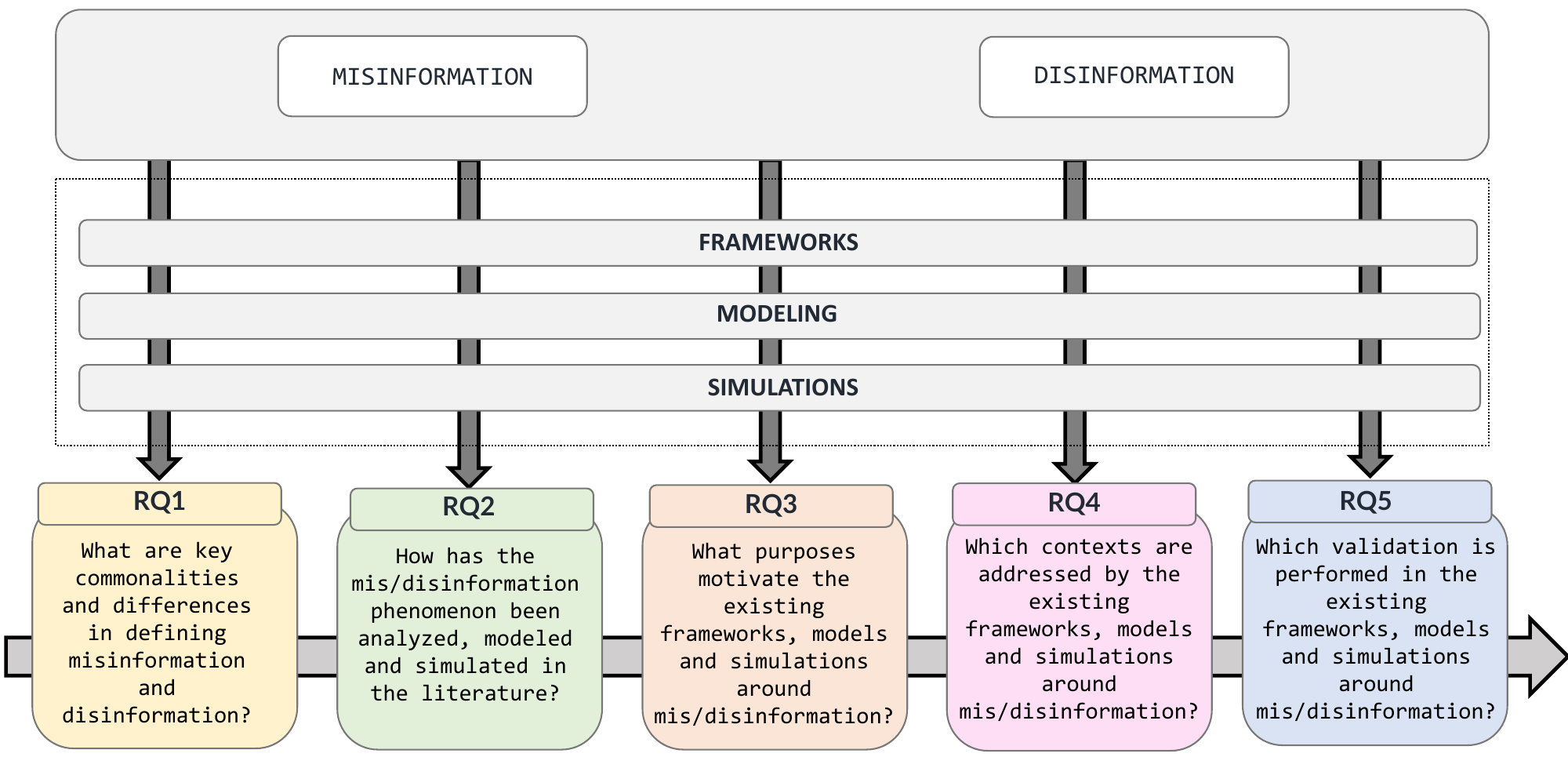}
\end{center}
\caption{Research questions related to frameworks, modeling and simulation of mis/disinformation.}
\label{fig:disinfo_process2}
\end{figure*}

This section presents a set of definitions to clarify the concepts of this systematic literature review. Figure \ref{fig:disinfo_process2} illustrates this study's key concepts and RQs. 

Initially, the concepts of misinformation and disinformation were explored within the mis/disinformation phenomena. There is not an established consensus on the definitions of both terms. Following the guidelines of the European Commission \cite{europaTacklingOnline}, in the present study, \textit{disinformation} is defined as ``false or misleading content spread to deceive or secure economic or political gain, which may cause public harm''. In contrast, \textit{misinformation} is ``false or misleading content shared without harmful intent, the effects can still be harmful''. 

Secondly, these two phenomena are analyzed through different means: \textit{frameworks} as conceptual abstractions that can be social or computational for studying or analyzing mis/disinformation phenomena, \textit{models} as logical or mathematical representations of any mis/disinformation dimension, and \textit{simulations} as experiments that emulate mis/disinformation dynamics. 

Specifically, five RQs were formulated to guide the investigation. The first RQ searched to clarify the definitions of misinformation and disinformation, given the absence of an official definition. This takes on great importance in the scientific field, where it is necessary to be very conscientious with the terms used. Subsequently, the study analyzed the methods used to represent, model, and simulate the mis/disinformation phenomenon. The objectives and purposes behind these approaches were also examined. Furthermore, the study investigated the contextual factors influencing these frameworks, models, and simulations. Finally, the validation methods employed were explored.


\subsection{Identification of research works}

The article collection was done on February 21, 2024, in two bibliographic databases: Scopus and the Web of Science (WoS), since they are the most widely used databases in different scientific fields \cite{DBLP:journals/corr/abs-1305-0377}. Scopus is the world's largest citation database of peer-reviewed research literature, with 27,950 active peer-reviewed journals \cite{elsevier2023}. Moreover, WOS is the second biggest bibliographic database, with almost 2.2 billion cited references from over 196 million records \cite{Wos2023}.

In order to perform the search on both databases, we restricted the query to title due to the high number of records. We used two pairs of terms to perform the query, one encompassing disinformation and misinformation. Therefore, papers were searched by title through the keywords ``\textit{disinformation}'' or ``\textit{misinformation}'', and ``\textit{framework}'' or ``\textit{simulat*}'' or ``\textit{model*}'' like the following:

\vspace{0.2cm}
\texttt{TITLE((misinformation OR disinformation) AND (framework OR model* OR simulat*))}

\vspace{0.1cm}

This query generated 303 initial studies (175 from Scopus and 128 from WoS).

\subsection{Screening of articles}

After obtaining the initial collection of 303 studies, 115 duplicated studies were identified and removed. With the collection of 188 studies, we formulated the following mandatory criteria to include a paper in the survey. The three authors conducted the screening process in consensus, applying the exclusion criteria sequentially so that a paper not matching a condition was removed.

\subsubsection{Exclusion by title and abstract}

\begin{itemize}
    \item The paper is not written in English or Spanish: three studies (1.6\% of the collection with 188 studies) were excluded.
    
    \item The paper is not published in conference proceedings, journals, or edited books/volumes (i.e., book chapters) to be peer-reviewed: four studies (2.13\%) were excluded.
    
    \item The paper does not contain a conceptual proposal related to mis/disinformation. Studies which not propose a framework, model, or simulation were excluded: 56 studies (29.8\%) of the papers were excluded.
    
    \item The paper does not address mis/disinformation in the context of digital media, including online platforms, social networks, the internet, or other digital media. Therefore, studies that addressed the phenomenon in the context of human memory and cognitive or clinical psychology were excluded: seven studies (3.72\%) of the papers were excluded.
\end{itemize}

The first screening excluded 72 works, with the remaining 116 studies to be evaluated next.

\subsubsection{Exclusion by full text}

\begin{itemize}
    \item The full text is open access, accessible through subscription, or sent by the authors upon request: two studies could not be obtained (1.72\% of the remaining 116 studies).
    
    \item The paper contains a conceptual proposal related to mis/disinformation: 54 studies (46.55\%) were excluded. One challenge we faced is that the new wave of AI produced a large volume of works in which the model is not inspected \cite{BONDIELLI201938}. Of the 54 studies excluded for not offering a conceptual proposal, 34 studies (62.96\% of the 54 papers) were presenting only applied ML. 
    
    \item The paper addresses mis/disinformation in the context of digital media, including online platforms, social networks, the internet, or other digital media: 3 studies (2.59\%) were excluded.
\end{itemize}

The second screening excluded 59 works, with the remaining 57 studies to include for review. The `BibTex' of the final collection with these studies is available online\footnote{\url{https://osf.io/58r2w}}.

\subsection{Inclusion of papers for review}

\begin{figure}[h!]
    \centering
    \includegraphics[width=0.6\columnwidth]{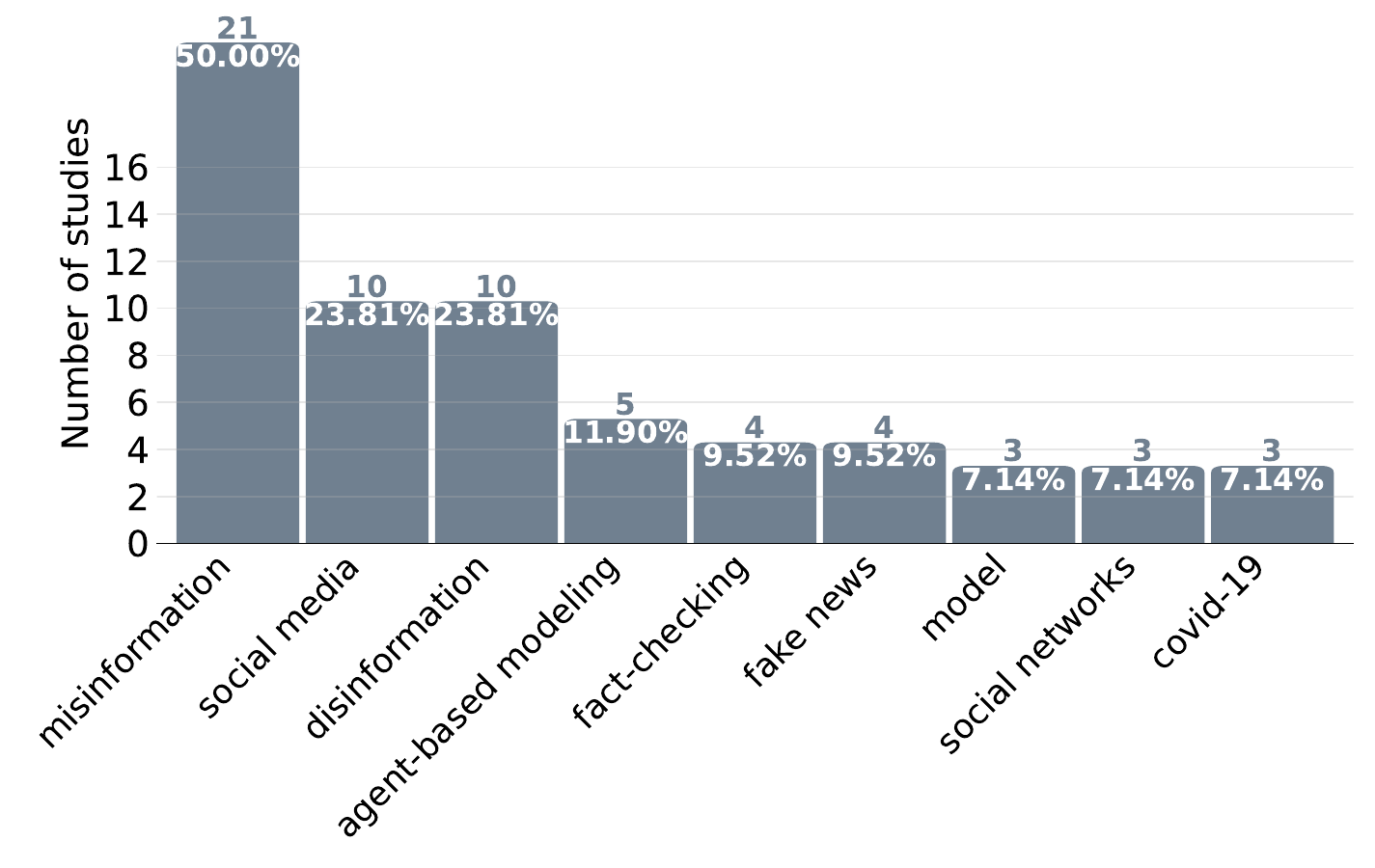}
    \caption{Distribution of keywords and rate across the studies included in the systematic review.}
    \label{fig:keywords}
\end{figure}

The keywords from the resulting 57 selected papers for the systematic review were collected.. The total sum is 211, while there are 149 unique keywords. The most frequent ones are presented in Figure \ref{fig:keywords}, having a high variability and strongly focused on misinformation (50\%), social media (23.81\%), and disinformation (23.81\%). Alternatively, the research topic covers other lines, such as agent-based modeling, fact-checking, fake news, models, social networks, and COVID-19.

\begin{figure}[h!]
    \centering
    \includegraphics[width=0.6\textwidth]{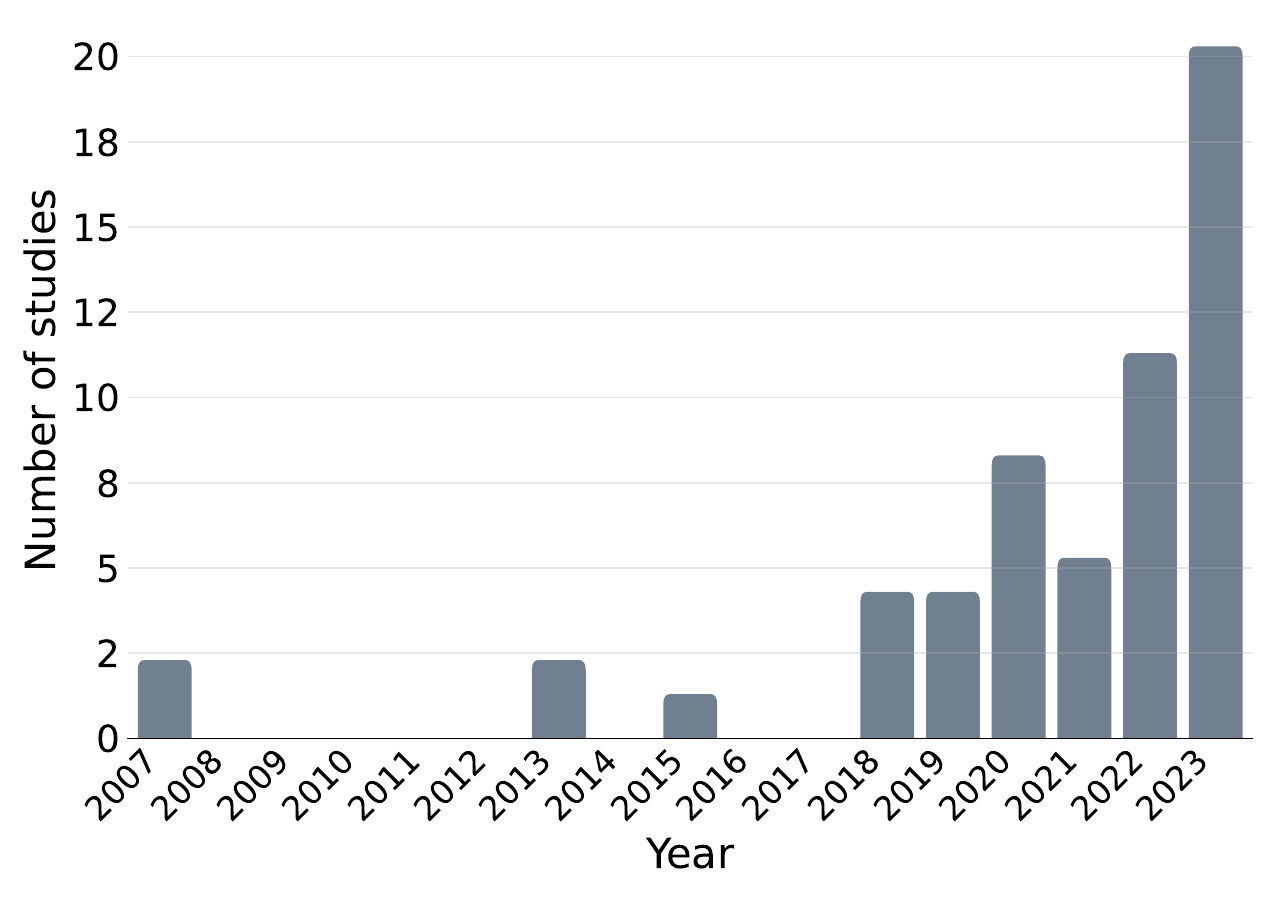}
    \caption{Number of selected studies and rate in the collection per year of publication.}
    \label{fig:years}
\end{figure}

On the other hand, Figure \ref{fig:years} shows the distribution of the 57 papers by publication year. Only five studies are found from 2007 to 2017 matching the review criteria. We see an increasing interest in this topic from 2018 onwards, having important peaks in 2020 (8 works) and 2022 (11 works). Finally, the increase in 2023 (20 works) aligns with the growth trend observed in recent years.

\subsection{Review and coding process}

The 57 articles were reviewed in detail during this phase, and relevant data were extracted from each study. Subsequently, a structured coding system was used to organize and categorize the information collected uniformly to answer the research questions in Section \ref{rqs}. It is worth noting that RQ1 is not coded due to its nature. In particular, an inductive coding scheme (also called open coding) was used for this coding process for RQ3, RQ4, and RQ5. This means that the codes created were based on the qualitative data itself \cite{emeraldQualitativeResearch}. This is an iterative process since researchers can add new codes, split an existing code into two, or compress two existing codes into one as they review data. Only a deductive coding scheme was used for RQ2, starting from three selected codes (\textit{Model}, \textit{Framework}, and \textit{Simulation}). Within each of these codes, further sub-codes were derived inductively. For all the RQs, the assigned codes are non-exclusive, i.e., one article can have several codes for one RQ. As a result, Table \ref{tab:codes} outlines the variable coding scheme we followed in this survey, indicating each RQ with its possible codes representing qualitative data.

\begin{table*}[ht!]
    \begin{tabular}{ P{1.4cm} c c c c}
    \hline
       \textbf{RQ2}  & \textbf{RQ3} &  \textbf{RQ4}&  \textbf{RQ5}
       \\ \hline
        \rule{0pt}{2.3ex}\rqtwo{Framework}  & \rqthree{Effectiveness evaluation} & \rqfour{Health} & \rqfive{Simulations with real-world data}\\
        \rule{0pt}{2.3ex}\rqtwo{Model} & \rqthree{ Mis/Disinformation detection} &  \rqfour{Politics} & \rqfive{Mathematical proof}\\
        \rule{0pt}{2.3ex}\rqtwo{Simulation} & \rqthree{Cognitive assessment} &  \rqfour{News articles} & \rqfive{Surveys}\\
        
         \rule{0pt}{2.3ex}& \rqthree{Mis/Disinformation conceptualization} & \rqfour{Defense} & \rqfive{Compared with real-world data}\\
         
         \rule{0pt}{2.5ex}&   &  \rqfour{Crises} & \rqfive{Focus groups}\\
         
         & &   \rqfour{E-commerce}  &\rule{0pt}{2.3ex}\\
        
        \hline 
    \end{tabular}
    \caption{The resulting variable coding scheme.}
    \label{tab:codes}
\end{table*}

All data generated from the screening and coding processes were in an Excel file, which is available online \footnote{\url{https://osf.io/e62px/}}.


\section{Results}\label{results}

\subsection{What are key commonalities and differences in defining misinformation and disinformation? (RQ1)}

In the analyzed studies, particularly in the literature from Anglo-Saxon countries, there is a plethora of terms and concepts that are used to refer to false, untrue, or half-true information, such as ``\textit{fake news}'' \cite{Govindankutty2023}, ``\textit{misinformation}'' \cite{Jiang2023}, or  ``\textit{disinformation}'' \cite{Bui2023}. Sometimes, these terms are used interchangeably \cite{disinfoTriangle}, which can be low discriminant. This RQ aims to uncover the key commonalities and differences that underlie its definitions, as well as disclose the number of papers that use the term `disinformation,' `misinformation,' or both terms and of the latter how many do so by giving definitions to discriminate between them making different use of each. Analyzing our collection of papers, the 57 papers, 37 employed the term misinformation, 17 used disinformation, and 3 employed both. However, 36 papers provide an explicit definition (17 articles explain the concept of misinformation, 11 studies define disinformation, and 8 works do it for both).

Regarding the definition of disinformation. By way of example, in \cite{KarinshakJin2023}, disinformation has been defined as false information designed to mislead. Moreover, in \cite{Kryshtanovych2023AnIM}, the phenomenon of disinformation is defined as a virus dangerous to the social organism capable of exponentially spreading in the information space in a short time, parasitizing primarily on the painful problems of the world as a whole, individual regions, and countries. Finally, D. Brody \cite{Brody2019modelingED} defined disinformation as modifying the information process.

In the case of misinformation, in \cite{doi:10.1177/10755470231207611} it is defined as false or misleading information that promotes claims already debunked by scientific evidence or expert opinions. Furthermore, in \cite{Zhu2023}, misinformation is used as an umbrella term to cover related concepts, including fake news, disinformation, false information, rumors, conspiracy theories, malignant information, inaccurate information, and more. Finally, in \cite{10.1145/2740908.2742572}, misinformation is defined as false claims that are mostly spread unintentionally. While, Safieddine \textit{et al.} \cite{8252201} defined misinformation as a piece of malicious information intended to cause undesirable effects on the general public, such as panic and misunderstanding, or to supplant valuable information and provide more studies that support that affirmation that is to say, i.e., it assigns the concept misinformation intentionality when spreading false information as in \cite{Zhou2007AnOM}. 

The relationship between both terms, disinformation and misinformation, is the consequences of both can be the same. For example, the authors in \cite{systems10020034} ``assume disinformation to be any deviant information that is intended to distort and mislead a target audience in a predetermined way'', argue as a key difference that disinformation is not only about the message itself but as a practice it has the potential to discredit the messenger and true information due to its close relationship with multiple social sectors, especially politics. Moreover, some papers defined misinformation and disinformation as terms that refer to information that does not directly reflect the `true' state of the world (e.g., distorted information or falsehoods). It does not differentiate between intentionality as in \cite{Tran2020MisinformationIC,osti_10275066}. On the other hand, 26 articles argue that the difference between disinformation and misinformation lies in the intentionality with which the information is disseminated. 

\begin{figure}[ht]
    \centering
    \includegraphics[width=0.7\columnwidth]{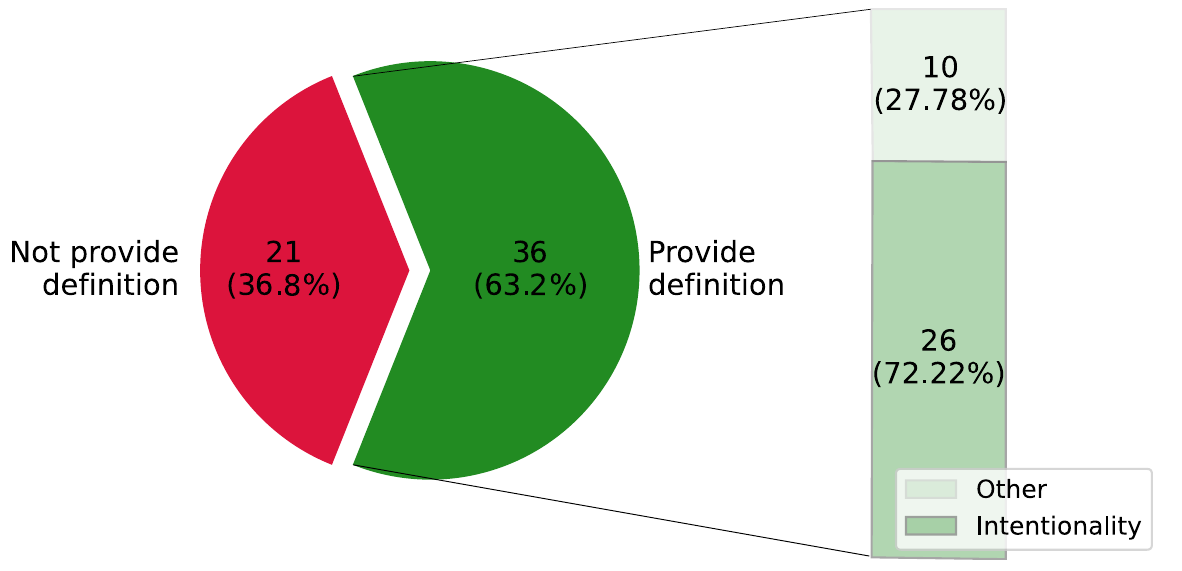}
    \caption{Paper distribution based on the definition provided.}
    \label{fig:rq1}
\end{figure}

In conclusion, this analysis reveals a consensus on the definitions of disinformation and misinformation. Misinformation refers to unintentionally false or inaccurate information, while disinformation entails deliberately spreading false or misleading information with the intent to deceive or manipulate. Figure \ref{fig:rq1} shows that 63.2\% of the papers provide a definition and 72.22\% of these papers highlight intentionality as a key concept in their definition. On the other hand, 27.78\% of these papers highlight any other aspect as a key concept in their definition, such as a virus, malicious information, or information that does not directly reflect the `true' or a modification of the information process.

\subsection {How has the mis/disinformation phenomenon been analyzed, modeled and simulated in the literature? (RQ2)}

In RQ2, our objective was to explore the existing literature on the representation of mis/disinformation. We coded the papers into three categories: \textit{frameworks}, \textit{models}, and \textit{simulations}, which will be divided into several subcategories:

\begin{itemize}
    \item \rqtwo{Frameworks}: Technical (9 studies, 52.94\%), Social (6 studies, 35.29\%), Technical and Social (2 studies, 11.77\%).

    \item \rqtwo{Models}: Epidemiology (18 studies, 41.86\%),  Opinion dynamics diffusion (7 studies, 16.28\%), Game-theoretic (6 studies, 13.95\%), Information diffusion (5 studies, 11.62\%),  Causal (2 studies, 4.65\%), Bayesian (2 studies, 4.65\%), Linguistic Pattern Recognition (1 study, 2.33\%), Ontologies (1 study, 2.33\%), Social learning (1 study, 2.33\%).

    \item \rqtwo{Simulations}: Belief updating (20 studies, 50\%), Countermeasures evaluation (10 studies, 25\%), Mis/Disinformation diffusion (9 studies, 22.5\%), Offensive evaluation (1 study, 2.5\%)
\end{itemize}

\subsubsection{Types of frameworks}

Frameworks serve as organizational structures and schemes of work that facilitate a holistic understanding of the mis/disinformation phenomena (such as actors, methods, means, consequences, types, etc). Our analysis reveals that there are three main types of frameworks:

\begin{enumerate}
    \item \rqtwo{Technical}: A technical framework in the context of mis/disinformation refers to a structured approach that focuses on utilizing computational and engineering methodologies to understand, analyze, and address the spread and impact of mis/disinformation. The primary objective of such frameworks is to develop tools, algorithms, and systems that can effectively analyze, detect, and combat mis/disinformation in digital information environments. 
    
    For example, the European project CALYPSO \cite{10.1007/978-3-031-05637-6_35} proposed the design of a framework that seeks to involve the audience in the early detection of mis/disinformation through a digital platform. The framework is based on gamification techniques to motivate users to participate in detecting and verifying mis/disinformation and in creating communities of practice involving different actors with different levels of responsibilities and privileges, such as readers, journalists, and experts. 
    
    Moreover, in \cite{ABOUZEID2022100341}, the authors proposed a technical framework that materialized in a modular implementation in Python that combines AI models and data visualization techniques to provide a practical resolution to emergency responders to help support their decision-making when working on an infodemic crisis.

    Furthermore, in \cite{10.1145/3581783.3612704}, the authors proposed a framework to provide real-time, comprehensive, and explainable detection measures for combating mis/disinformation in the era of Generative AI Models (GenAI). The framework operates on four levels: signal, perceptual, semantic, and human, aiming to detect manipulation traces in AI-generated content and address technical flaws, misleading content, and propagation processes contributing to the spread of mis/disinformation.

    \item \rqtwo{Social}: These frameworks operate on the premise that combating mis/disinformation requires more than just technical solutions; it necessitates a comprehensive understanding of societal dynamics, human behavior, and the intricate networks through which information propagates. Social frameworks seek to manage mis/disinformation from a broader perspective, recognizing that the challenge extends beyond technological facets.
    
    By way of example, in \cite{Karlova2013ASD}, Karlova and Fisher proposed a social diffusion framework for illustrating how information, misinformation, and disinformation are formed, disseminated, judged, and used in terms of key elements, beginning with milieux.
    
    Moreover, the authors in \cite{disinfoTriangle} proposed a framework called ``the disinformation and misinformation triangle'' to show the three determinants of disinformation and information proliferation in digital news: the pathogen, host, and environment.

    Furthermore, in \cite{Amazeen2023}, the authors proposed a framework called ``Misinformation Recognition and Response Model (MRRM),'' which explores how individuals recognize and respond to disinformation. It emphasizes individual factors, situational factors, motivational goals, and message characteristics that influence information processing. The framework also highlights the importance of prebunking, source credibility, message format, and cognitive coping strategies in dealing with disinformation.

    Finally, in \cite{KarinshakJin2023}, the authors proposed a framework that focuses on understanding the effects of AI on disinformation attacks and counterattacks and how organizations can proactively manage and respond to disinformation. It emphasizes the importance of a systematic and coordinated approach to counter disinformation efforts. The framework includes strategies such as detection and monitoring, corrective communication, flagging and removing content, and promoting media literacy. Additionally, the framework differentiates between malicious and non-malicious human and AI actors in digital environments. 

    \item \rqtwo{Technical and social}: These hybrid frameworks bridge the gap between human-centered thinking and innovative technical solutions, recognizing that an effective response to disinformation requires a multidisciplinary strategy.
    
    For example, the framework DISARM \cite{Terp2022} is the open-source, master framework for fighting disinformation through sharing data and analysis and coordinating effective action. The framework has been developed, drawing on global cybersecurity best practices. But it is not only technical, it is used to help communicators, from whichever discipline or sector, to gain a clear shared understanding of disinformation incidents and to immediately identify defensive and mitigation actions that are available to them.

    Moreover, the disinformation threat framework proposed in 2023 by \cite{EURECOM+7114} characterizes the disinformation threat across domains by mapping out potential threat actors, their motives and capabilities, their observed patterns of attack, the attack channels they use, and the audiences they target, considering the attacker side, their tactics and approaches.
    
\end{enumerate}

In our analysis, there are 17 studies (26.32\%) that proposed a framework. 
The social framework is the most used (9 studies, 52.94\%), followed by technical frameworks (6 studies, 35.29\%). At the same time, only two studies (11.77\%) used both approaches.


\subsubsection{Types of models}

To understand the phenomenon of disinformation, having structures to represent its different elements is a key aspect. However, these models can be conceptual and supported by robust mathematical models to enhance their validity and analytical capabilities. From our analysis, ten primary categories are found:

\begin{enumerate}
    \item \rqtwo{Epidemiology}: As infectious diseases spread in populations, disinformation can spread rapidly, affecting public opinion, attitudes, and, ultimately, people's behavior. The simile between an epidemic and disinformation is that exposure to false information can influence the adoption of erroneous beliefs, generating a kind of social ``contagion.'' Changes in people's susceptibility to disinformation and their susceptibility to corrective actions can be modeled like that of acquired immunity in epidemiology, adapting to the disinformation phenomenon, which an individual can adopt different roles.
    
    For example, the authors in \cite{panchal2022mathematical} proposed a novel epidemiology model called ``SEHIR'' with an additional compartment of people not reacting instantly or at all to the disinformation called hibernators, to obtain more clarity on the pattern of spread of fake news. Moreover, in \cite{Gavric_2019}, the authors proposed a ``fuzzy'' version from the epidemiology model SI (Susceptible - Infected), incorporating fuzzy parameters and factors such as user influence, the likelihood of sharing disinformation and the strength of association between different users in the social network to more accurately describe the spread of disinformation in the social network.
    
    Other epidemiology models such as SIR (Susceptible-Infected-Recovered) were used, for instance, in \cite{10.1057/s41599-023-01998-z} where authors employed the SIR model as a basis, to understand how disinformation spreads in an epidemic-like manner, with misinformed individuals seeking to affect a susceptible population by transmitting messages with false information including specific elements of the disinformation diplomacy strategy, such as organic and paid reach, bots and trolls, providing a clearer picture of how the elements of disinformation interact to affect a susceptible population.
    
    As a last example, the authors in \cite{9260064} used an existing epidemiology model called ``SBFC,'' which has three states (Susceptible - Believer - Fact Checker), which allows them to study how fake news spreads and is countered in a population of agents, who can decide whether to believe or check the veracity of the information they receive.

    \item \rqtwo{Information diffusion}: Information diffusion models provide conceptual and mathematical tools for understanding how ideas, information, and narratives spread through social networks and society, influencing public opinion and individual and collective decision-making, representing the flow of information between individuals or nodes in a network, and considering factors such as the speed of information transmission, or the susceptibility of individuals to accept or reject false information. These models help to comprehend the mechanisms underlying the spread of disinformation, identify key factors that contribute to its proliferation, and develop targeted interventions or strategies to curb its impact and promote accurate information dissemination in society.
    
    By way of example, in \cite{ABOUZEID2022100341}, the authors proposed a new information diffusion model composed of two models. One of them is responsible for predicting the behavior of individual users on the social network based on a Multivariate Hawkes process (MHP), which models the occurrence of temporal or spatiotemporal asynchronous events by capturing their mutual dependencies. The other introduces additional data to the network to alter the diffusion's future outcomes.
    
    On the other hand, in \cite{9381446}, the authors used the existing multicampaign independent cascade model (MCICM) introduced in 2011. The MCICM simulates the spread of true and false information through a network's positive and negative seed nodes. It is used to study how false information spreads and how to counteract its impact, being able to identify key nodes to disseminate correct information and combat disinformation.
    
    Finally, the authors in \cite{LIN2023105312} adapted the drift-diffusion model (DDM) to model the decision-making process. The model assumes a decision-making process in which an initial bias in favor of one discrete option over another changes gradually until a decision boundary is reached, i.e., a critical point in the model where a discrete decision is made based on the accumulation of evidence over time. This model helps illuminate why precision prompts (signals or information provided to individuals to help them discern the integrity of the information they are receiving) can decrease the spread of disinformation by examining how people think more or differently when given these prompts.

    \item \rqtwo{Opinion dynamics}: These models draw on social network theory and social psychology to explore how opinions are transmitted and modified over time, how social interactions can shape individual beliefs and attitudes, and how these dynamics affect the collective perception of a topic. These models consider factors such as social influence, confirmation bias, cognitive biases, and information exposure to understand the evolution of opinions toward disinformation. In the context of disinformation, where false information can spread rapidly, understanding how opinions spread and persist is a research topic.
    
    For example, in \cite{10.5555/3398761.3399046}, the authors proposed a new opinion dynamics model to discuss how a disinformation campaign can change people's minds. Modeling the interaction between conspirators who spread false information and inoculators who attempt to protect the population susceptible to being influenced by such disinformation. Something similar is done in \cite{Brody2019modelingED}, where the authors analyze with a novel opinion dynamics model the impact of disinformation on the results of a political election, introducing disinformation as an additional term in the information process.

    \item \rqtwo{Game-theoretic}: Game theory is a multidisciplinary field that originated in economics and mathematics. It focuses on studying rational agents' strategic decisions to maximize their goals in an interactive environment. This modeling is conducted by formulating a game or strategic interaction between agents within a system or network. In the case of disinformation, game theoretic models allow us to examine how different actors make decisions informed by incentives and consequences.
    
    For instance, in \cite{10002219}, the authors proposed a game theoretic model consisting of a repeated series of adapted impartial games from the original Nim game, played by a reinforcement learning agent (Q-learning), to learn the `truth' modeled by the optimal Nim strategy. Another example is the approach in \cite{10.1155/2022/1136144}, where the authors develop a novel tripartite evolutionary game consisting of three players, including network media, government, and netizens, and analyze the mutual influences of the player.

    \item \rqtwo{Causal}: Causal models represent causal relationships within an individual system or population. They facilitate inferences about causal relationships from statistical data. This modeling analyzes the cause-and-effect relationships between several variables. In the context of disinformation can help to understand the cause-and-effect relationships underlying the generation, propagation, and counteraction of disinformation in diverse information environments.
    
    For example, in \cite{systems10020034}, the authors proposed a causal model adapted from the epidemiology model SIR (Susceptible - Infected - Recovered) to model and simulate scenarios of disinformation propagation in social networks caused by bots. Moreover, the authors in \cite{doi:10.1080/01292986.2020.1811737} examined how cognitive, affective, and environmental factors interplay to affect the acquisition and diffusion of food safety disinformation by analyzing a national sample of Chinese Internet users using surveys and proposed a novel causal model called ``emotion-driven cognitive dissonance model'' to explain how disinformation about food safety is diffused both online and offline and provide recommendations to help battle it.

    \item \rqtwo{Bayesian}: Bayesian inference techniques refer to statistical methods and algorithms that update beliefs or make predictions based on Bayesian principles. These techniques are pivotal in Bayesian models, which allow the modeling of the spread of disinformation from a probabilistic perspective, considering the uncertainty associated with how individuals perceive and interpret information. These models allow for prior knowledge, such as on the reliability of sources or the susceptibility of certain groups to disinformation.
    
    For instance, in \cite{Cohen_2020}, Cohen \textit{et al.} proposed a Bayesian multi-agent model for social networks. This model is based on a partially observable Markov decision process (POMDP). This mathematical model of sequential decision-making where complete information about the system's current state is unavailable. The model proposed in this study considers three main factors: the similarity of the user with the evaluator, the credibility of the evaluator, and the actual ratings provided.

    Moreover, in \cite{doi:10.1027/1016-9040/a000498} Zmigrod \textit{et al.} built a Bayesian model of cognition and conceptualizes political disinformation receptivity as a cognitive inference problem where the reliability of incoming disinformation is weighed against the reliability of prior beliefs. By modeling the integration of prior beliefs and new information, the model aims to explain when individuals adopt or reject political disinformation and how they creatively generate interpretations rather than passively discern truth versus falsehood.

    \item \rqtwo{Linguistic Pattern Recognition}: A Linguistic Pattern Recognition model is a computational approach that identifies and analyzes linguistic patterns in textual data. These models focus on recognizing specific linguistic and syntactic features that may indicate the presence of inaccurate or misleading information. The purpose of using these models in the context of disinformation is to automate the detection process, enhance accuracy in identifying deceptive content, and provide insights into the linguistic features associated with disinformation.
    
    For example, in \cite{4161614}, a new model is proposed to use non-parametric statistical methods to identify linguistic patterns in text data and compile ``hits'' and scores for article content that is directly related to the indicators of multiple computational social science models.

    \item \rqtwo{Ontologies}: In computer science and communication science, an ontology is a formal definition of types, properties, and relationships between entities that actually or fundamentally exist for a particular domain of discourse. It is a practical application of philosophical ontology with a taxonomy. In disinformation, ontologies can be designed to represent the structure of information and the relationships between entities of interest. This modeling approach can help to categorize different types of deceptive content, identify patterns of disinformation propagation, and analyze the impact of disinformation on various stakeholders. Moreover, ontologies facilitate the integration of diverse data sources and developing automated systems for disinformation detection, mitigation, and response. 
    
    Motivated by the importance of ontology support in information sharing and integration, in \cite{Zhou2007AnOM}, the authors made the first effort toward building an ontology-supported disinformation model. They employed ontology, in general, and W3C standards\footnote{\url{https://www.w3.org/TR/owl-features}}, in particular, to represent disinformation. This model enhances disinformation detection while laying the theoretical foundation for building a digital disinformation library that can foster disinformation sharing in the research community. This model promotes interdisciplinary collaboration by linking information and social science theories, artificial intelligence, systems analysis, and design.

    \item \rqtwo{Social learning}: Social learning modeling is constructing computational models of how individuals acquire knowledge, beliefs, and behaviors through social interactions within a community or society, considering social factors such as peer influence, homophily, and network structure. These models provide a broader, empirically grounded, and analytically tractable framework for understanding information aggregation. However, their main drawback is that they tend to create a long-term population consensus and do not account for opinion heterogeneity or polarization. Polarization can be generated by introducing `stubborn' agents who remain fully committed to their original views rather than interacting and learning from their neighbors, a mechanism reminiscent of confirmation bias. 

    To capture the effect of large-scale confirmation bias on social learning, the authors in \cite{38729e78dce24d22b6cf06e4b13384b5} proposed a novel social learning model where most participants in a network update their beliefs unbiasedly based on new information. In contrast, some participants reject information incongruent with their preexisting beliefs.
    
\end{enumerate}

\vspace{0.5cm}
In total, 43 studies are identified that proposed a model. Only one study was identified that made use of two types of the above-described models at the same time. In \cite{10.1109/GLOBECOM46510.2021.9685197}, the authors proposed a game-theoretic opinion model for the subjective opinions and behavioral strategies of attackers, users, and defenders. Further, they investigated which opinion model(s) can better help combat disinformation (i.e., not believing disinformation). Moreover, we distinguish when a proposal is completely new and novel (\textit{New proposal}), when a proposal already existing in the literature is reused (\textit{Existing proposal}), and when an existing proposal is adapted (\textit{Adapted proposal}).

\begin{figure*}[ht]
    \centering
    \includegraphics[width=0.85\textwidth]{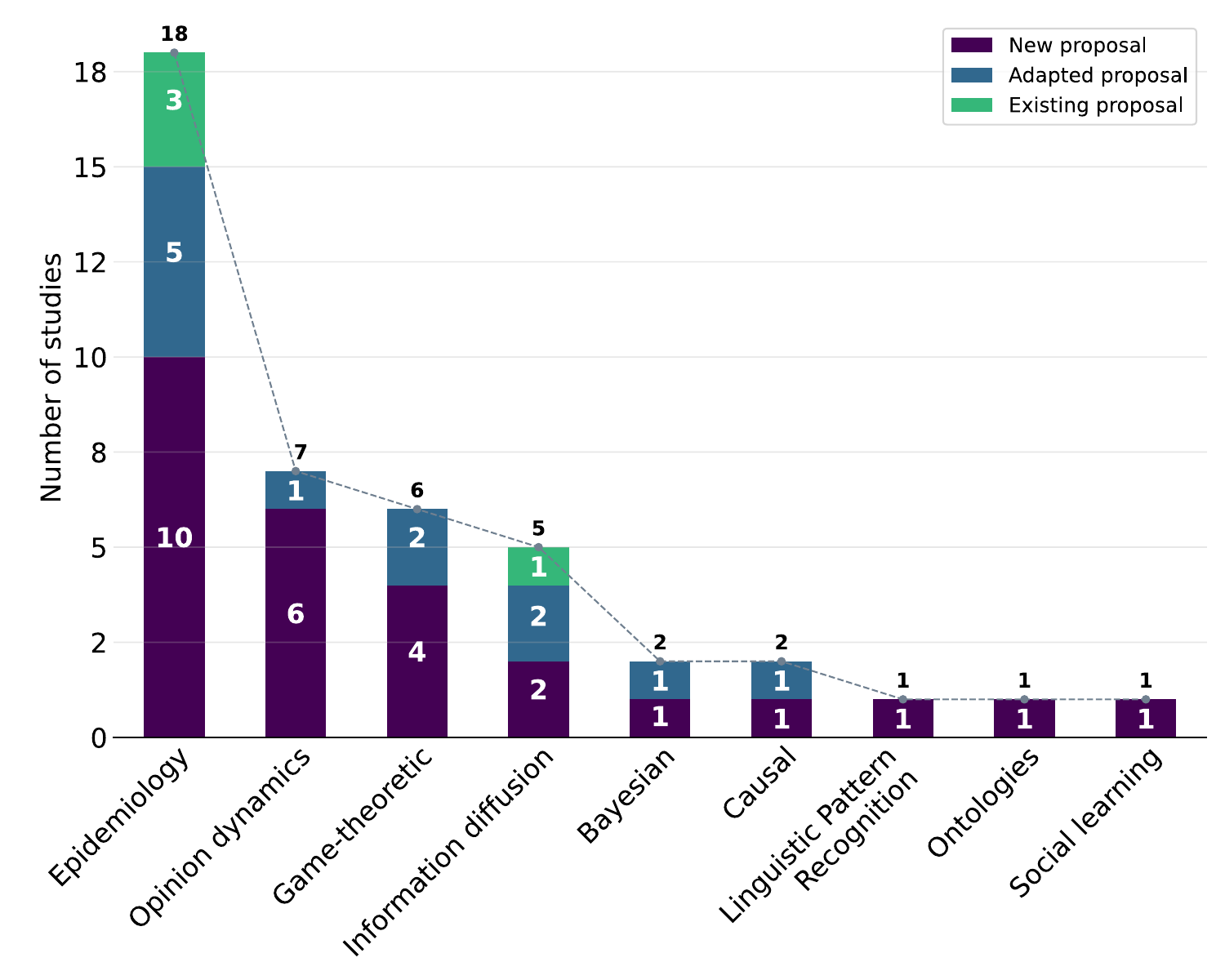}
    \caption{Paper distribution by model type and novelty of the proposal.}
    \label{fig:rq2-m}
\end{figure*}

In Figure \ref{fig:rq2-m}, we can see that \rqtwo{epidemiology} models are the most used in the literature, simulating its propagation similar to the spread of diseases in a population (18 studies, 41.86\% of studies that proposed a model). Most of these studies are novel proposals incorporating more or different states an agent can transit. The \rqtwo{opinion dynamics} models are the second most used (7 studies, 16.28\%). In this case, except one, all studies proposed novel models for representing and analyzing disinformation. Following are the \rqtwo{game-theoretic} models (6 studies, 13.95\%) and \rqtwo{information diffusion} models (5 studies, 11.62\%). The rest of the models are not extensively used, with two types of models used in two studies and three types of models in only one study.

\subsubsection{Types of simulations}

Simulations offer a dynamic perspective on spreading mis/disinformation in controlled environments and provide a valuable tool for evaluating mis/disinformation models. From the coding process, we present four categories:

\begin{enumerate}

    \item \rqtwo{Belief updating}: Belief updating simulation is a type of simulation employed to examine the dynamics of belief alteration among individuals following exposure to mis/disinformation. This type of simulation is used to study how people change their beliefs when exposed to mis/disinformation. These simulations aim to determine how misleading information affects people's beliefs. This type of simulation can be used to study various factors, such as the credibility of the information source, the presence of cognitive biases, and the participants' prior knowledge level.
    
    For example, in \cite{10.1145/2740908.2742572}, the authors simulated their stochastic epidemic model to analyze the transition between states (susceptible, believer, and fact-checker) when mis/disinformation and its relative debunking were spread in a social network.
    
    On the other hand, in \cite{9260064}, the authors employed ``NetLogo'', one of the most relevant simulation platforms, to simulate the diffusion of a piece of mis/disinformation. In particular, they used the ``SBFC'' model that describes the spread of misinformation as a competition between a fake news story and its debunking in a population of agents. Agents can switch between being Susceptible, Believers, or Fact-checkers, reflecting how individuals' beliefs evolve depending on the information they receive from their neighbors and their environment.

    \item \rqtwo{Countermeasures evaluation}: Countermeasures evaluation simulations aim to analyze how corrective or preventive measures can affect the spread and perception of mis/disinformation in specific information environments, exploring hypothetical scenarios and assessing the impact of various countermeasures. This makes it possible to identify more effective strategies to combat mis/disinformation and improve society's resilience in the face of this phenomenon.
    
    For instance, the authors in \cite{Bui2023} developed an agent-based model in Python to simulate and investigate the diffusion of prebunking interventions through three different stereotypical mis/disinformation campaign attack scenarios from a macro-level perspective and how prebunking intervention can help in the combat with mis/disinformation.
    
    On the other hand, in \cite{8252201, Pourghomi2018TheSO}, the authors used the third-party software called ``Biolayout'' as a three-dimensional modeling tool and a platform for simulating different scenarios. They propose the implementation of a ``right-click authenticate'' button as a countermeasure to combat the spread of mis/disinformation online. This button would allow users to right-click on a news item, image or video to perform real-time verification of its origin, original metadata, observations cited by editors, and crowd-sourced feedback. 

    Moreover, the authors in \cite{ButtsBollmanMurillo2023} simulated their model to examine and optimize strategies for combating mis/disinformation in social networks, such as content moderation, education, and counter-campaigns. Concluding that the most promising strategies for combating mis/disinformation are education-based policies that increase skepticism and counter-campaign policies.

    Furthermore, the authors in \cite{pan-etal-2023-risk} simulated mis/disinformation using LLMs. Specifically, they used GPT-3.5 as the generator. Countermeasures against mis/disinformation have been simulated through three defense strategies: prompting, detection, and majority voting. These strategies were evaluated to mitigate the harm caused by LLM-generated mis/disinformation in Open-Domain question-answering (ODQA) systems.

    \item \rqtwo{Mis/Disinformation diffusion}: Mis/Disinformation diffusion simulations replicate and analyze the spread of false or misleading information throughout interconnected networks, such as social networks, information networks, or other graph structures. These simulations consider user interactions, content-sharing mechanisms, network topology, and information propagation pathways. The purpose of conducting these simulations is to gain insights into the mechanisms and patterns of mis/disinformation spread, understand the dynamics of information flow within networks, identify influential nodes or factors that accelerate or hinder dissemination and develop strategies to mitigate the harmful effects of mis/disinformation.
    
    For example, the authors in \cite{systems10020034} used the third-party software called ``Stella Architect'' to simulate their causal model, using system dynamics as the main technique, in scenarios of mis/disinformation propagation in social networks caused by bots creating and sharing content, while modifying parameters, such as the rate of activation and deactivation of bots, to assess their impact on the spread of mis/disinformation on social networks caused by bots.
    
    In addition, in \cite{4161614}, the authors simulated their TRAQ-M (Tracking Analysis, Quantification-Mitigation) framework, which is a platform with a computational system that applies Social Science Models and non-parametric statistical methods to understand complex human behavior patterns through language using Language Pattern Recognition models. These simulations focus on extracting linguistic patterns to compile reliable information and detect possible sources of misinformation in the open press.

    Furthermore, in \cite{doi:10.1142/S0217979224502837}, the authors simulated their proposed new model called Susceptible, Exposed, Negative-Infected, Positive-Infected, Recovered Population (SENPR) to describe the dynamics of mis/disinformation in OSNs, analyzing the factors influencing its spread, and suggesting management strategies to suppress false information in emergencies. The model was simulated using the third-party software ``AnyLogic'' to build the multi-agent simulation model of the SENPR model and construct the network propagation model.

    \item \rqtwo{Offensive evaluation}: As simulations can serve as a valuable tool to evaluate the impact of countermeasures, they can likewise serve to evaluate attacks of mis/disinformation that can consist of introducing fake or misleading content in a simulated information environment. Offensive evaluations are designed to assess the effectiveness and impact of these offensive strategies, such as coordinated campaigns or targeted messaging strategies. Simulating offensive tactics makes it possible to evaluate the potential risks posed by such strategies, identify vulnerabilities in simulated information dissemination systems, and develop countermeasures to mitigate the harmful effects of offensive mis/disinformation campaigns.
    
    Only one study was identified in this category, in \cite{9004942} Beskow \textit{et al.} employed bots to simulate distinct mis/disinformation attacks on Twitter. In the ``\textit{Backing}'' attack, bots are programmed to bolster key influencers within a network, amplifying their messages and expanding their reach. Alternatively, in the ``\textit{Bridging}'' attack, bots are configured to support various communities, facilitating connections and the dissemination of information across these diverse groups. 
    
\end{enumerate}

As a result, 40 studies that proposed simulations are identified. Of these studies, three were identified that made use of two types of the above-described simulation types at the same time. By way of example, in \cite{10002219}, the dissemination of mis/disinformation and how it affects people's opinions is simulated using a game-theoretic model previously defined. Moreover, in \cite{echochambers}, the authors simulated the spread of mis/disinformation and the opinion updating to study a possible relationship between echo chambers and the viral spread of mis/disinformation. Lastly, in \cite{6637343}, the authors simulated the belief change in social networks through models based on evolutionary game theory and evolutionary graph theory. At the same time, extensive simulations are performed using real social network datasets to study the process of online information diffusion.

\begin{figure*}[t]
    \centering
\includegraphics[width=0.85\textwidth]{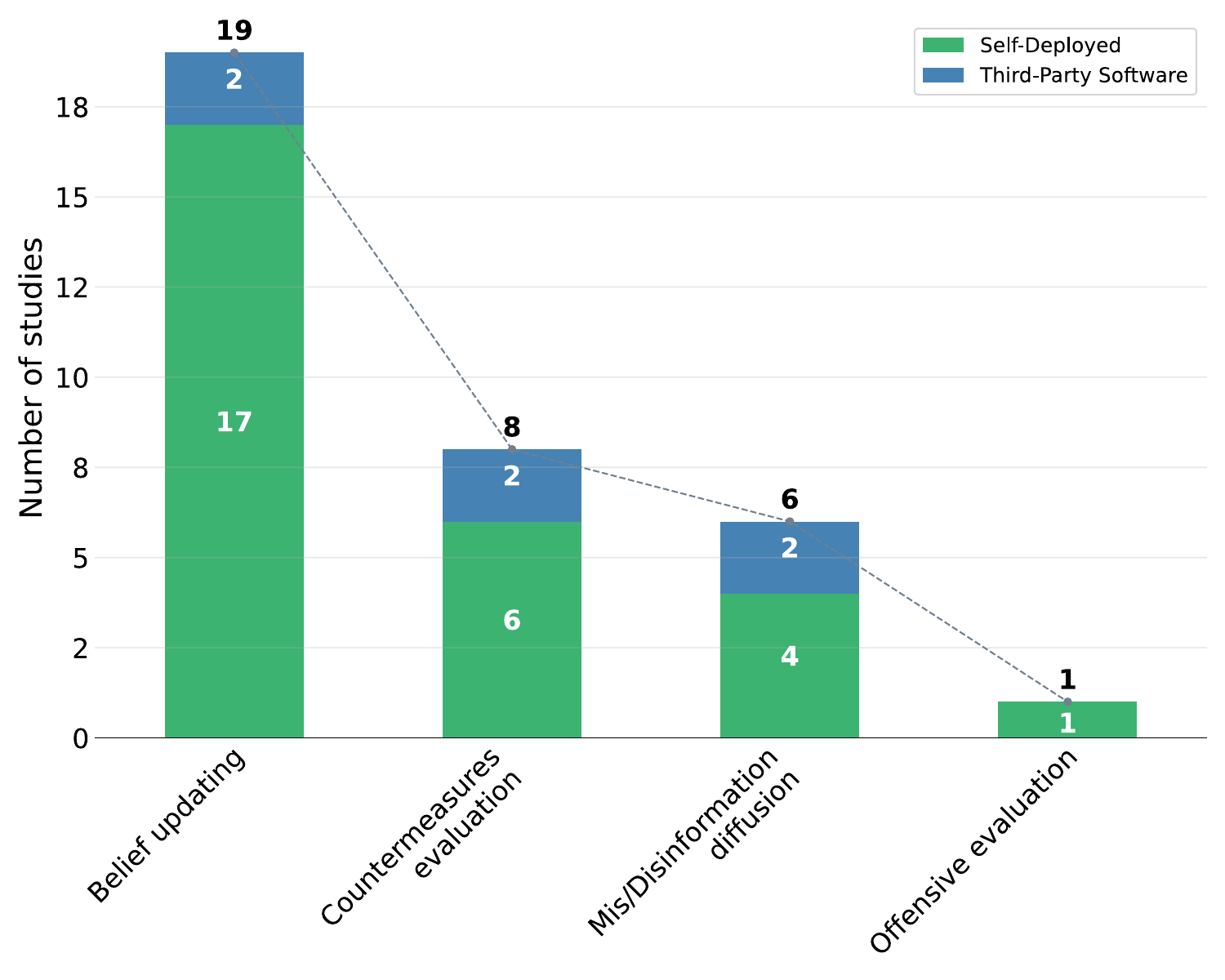}
    \caption{Paper distribution by type of simulation and development of the proposal.}
    \label{fig:rq2-s}
\end{figure*}

Similar to what has been done in the models, we will distinguish between simulations conducted autonomously by researchers, such as programming their own simulations (\textit{Self-developed}), and simulations that utilize existing third-party software (\textit{Third-party software}). Figure \ref{fig:rq2-s} illustrates that the most used type of simulations is \rqtwo{Belief updating} (20 studies, 50\% of studies that proposed simulations). Of these simulations, the majority (18 studies, 90\% of belief updating simulations) were self-developed by the authors. The second type of most used simulations was \rqtwo{Countermeasures evaluation} (10 studies, 25\%) equal to the previous type. The authors self-developed the majority (8 studies, 10\% of countermeasures evaluation simulations). Following this is the \rqtwo{Mis/Disinformation diffusion} (9 studies, 22.5\%) and \rqtwo{Offensive evaluation} (1 study, 2.5\%).

\subsubsection{Relationship between frameworks, models and simulations}

To analyze the relationship between frameworks, models and simulations, we employ a log-linear analysis to compare the three dimensions to determine if there is an association between them. Table \ref{tab:log_linear} shows the cell count of a log-linear analysis and the most typical associations in percentage. The reviewed studies tend to propose models together with simulations (50.88\%). Then, it is typical for the framework proposals (24.56\%) without modeling or simulation. Finally, studies propose unique models (12.28\%) and unique simulations (7.02\%) or jointly present a framework with modeling and simulation (5.26\%). 

As a result, the $z$-score values from the log-linear analysis reveal significant relationships among the use of frameworks, models and simulations. While, $p$ represents the probability value associated with the observed relationship in the log-linear analysis, indicating the significance level of the result. Notably, a substantial negative association is observed between using a framework and the presence of a model or simulation ($z$ = -3.362, $p$ = 0.001), indicating that articles proposing frameworks do not usually incorporate modeling or simulations. This fact makes sense due to a framework's conceptual or theoretical nature. Conversely, a positive and significant relationship is identified between the use of models and simulations ($z$ = 1.771, $p$ = 0.077), indicating that simulations precise a certain modeling or validate a particular model.

Once the relationship between the presence or absence of a proposal related to a framework, simulation, or model has been analyzed, we will examine whether there is a connection between the specific type of each proposal. Table \ref{tab:proposals} has been created to illustrate the number of combinations present for each type of proposal that appears more than once. We can see that 11 articles combine epidemiology models with belief updating simulation (19.3\%). Then, 9 studies proposed a social framework without a model or simulation (15.79\%). As expected, 4 opinion dynamics models are validated with belief updating simulations (7.02\%). Finally, it was found that sometimes simulation of countermeasures is not based on particular models or frameworks (4 studies, 7.02\%), but in other situations, it is employed through information diffusion models (3 articles, 5.26\%).

\begin{table*}[t!]
    \centering

    \begin{tabular}{p{6cm} P{1.7cm} P{1cm} P{1.5cm} P{0.8cm} P{1.5cm}}
    \hline
        \centering\textbf{Articles} & \textbf{Framework}& \textbf{Model} &  \textbf{Simulation}&   \textbf{Total} & \textbf{Ratio (\%)}\\
         \hline
         \cite{Bui2023,Jiang2023,DOUVEN2021103415,Govindankutty2023,10.1109/GLOBECOM46510.2021.9685197,38729e78dce24d22b6cf06e4b13384b5,10002219,Liu2022,systems10020034,LIN2023105312,10.1155/2022/1136144,osti_10275066,9381446,10.5555/3398761.3399046,Brody2019modelingED,echochambers,Tambuscio2001,9004942,10.4018/IJSWIS.300827,panchal2022mathematical,6637343,9260064,10.1145/2740908.2742572,Cohen_2020,JIANG2021125572,10.1057/s41599-023-01998-z,ButtsBollmanMurillo2023,GhoshDas2023,doi:10.1142/S0217979224502837} &  \xmark & \greencheck &  \greencheck  & \vspace{0.01cm} 29 & \vspace{0.01cm} 50.88\%\\
         \hline
         \cite{Kryshtanovych2023AnIM,10083447,Agarwal2021,10.1007/978-3-031-05637-6_35,Tran2020MisinformationIC,Terp2022,Karlova2013ASD,disinfoTriangle,EURECOM+7114,10.1145/3603163.3609057,KarinshakJin2023,10.1145/3581783.3612704,Zhu2023,Amazeen2023} &\greencheck  & \xmark &  \xmark &  \vspace{0.001cm} 14 & \vspace{0.005cm} 24.56\%\\
         \hline
         \cite{doi:10.1080/01292986.2020.1811737,Zhou2007AnOM,Gavric_2019,10.1007/978-3-031-35927-9_28,10.14722/ndss.2023.24293,doi:10.1027/1016-9040/a000498,doi:10.1177/10755470231207611} &\xmark & \greencheck & \xmark & \vspace{0.005cm}7 & \vspace{0.005cm}12.28\%\\
         \hline
         \cite{Kirk2020,8252201,pan-etal-2023-risk,Pourghomi2018TheSO} &\xmark & \xmark & \greencheck & \vspace{0.005cm}4 & \vspace{0.005cm}7.02\%\\    
         \hline
         \cite{9920336,ABOUZEID2022100341,4161614} &\greencheck & \greencheck & \greencheck & \vspace{0.005cm}3 & \vspace{0.005cm}5.26\%\\
    \hline
     & & & & 57 & 100\%\\
    \hline
    \end{tabular}
        \caption{Proposal of frameworks, models and simulations in mis/disinformation studies.}
    \label{tab:log_linear}
\end{table*}

\begin{table*}[h!]
    \centering
    \label{tab:proposals}
    \begin{tabular}{p{2.3cm}P{2cm}P{3cm}P{3.6cm}P{1cm}P{1cm}}
    \hline
        \centering\textbf{Articles} &  \textbf{Framework} & \textbf{Model} & \textbf{Simulation} &  \textbf{Total} & \textbf{Ratio}\\
         \hline
         \multirow{3}{=}{\centering\cite{Jiang2023,Govindankutty2023,Liu2022,osti_10275066,Tambuscio2001,10.4018/IJSWIS.300827,panchal2022mathematical,9260064,10.1145/2740908.2742572,JIANG2021125572,echochambers}} &  \multirow{3}{*}{\centering\xmark} & \multirow{3}{*}{Epidemiology}   &  \multirow{3}{*}{Belief updating} &  \multirow{3}{*}{11} &  \multirow{3}{*}{19.3\%}\\
         & & & & & \\
& & & & & \\
         \hline
         \multirow{3}{=}{\centering\cite{Kryshtanovych2023AnIM,10083447,Agarwal2021,Tran2020MisinformationIC,Karlova2013ASD,disinfoTriangle,KarinshakJin2023,Zhu2023,Amazeen2023}} & \multirow{3}{*}{Social} & \multirow{3}{*}{\xmark} &   \multirow{3}{*}{\xmark}   &  \multirow{3}{*}{9} &  \multirow{3}{*}{15.79\%}\\
         & & & & & \\
         & & & & & \\
         \hline
         \multirow{2}{=}{\centering\cite{Kirk2020,8252201,pan-etal-2023-risk,Pourghomi2018TheSO}} &  \multirow{2}{*}{\centering\xmark} & \multirow{2}{*}{\centering\xmark} & \multirow{2}{*}{Countermeasures evaluation} &  \multirow{2}{*}{4} &  \multirow{2}{*}{7.02\%}\\
         & & & & & \\
         \hline
         \multirow{2}{=}{\centering\cite{DOUVEN2021103415,10.5555/3398761.3399046,Brody2019modelingED,10.1109/GLOBECOM46510.2021.9685197}} & \multirow{2}{*}{\centering\xmark} & \multirow{2}{*}{Opinion dynamics} & \multirow{2}{*}{Belief updating} & \multirow{2}{*}{4} & \multirow{2}{*}{7.02\%}\\
         & & & & & \\
         \hline
         \multirow{2}{=}{\centering\cite{echochambers,10.1057/s41599-023-01998-z,GhoshDas2023,doi:10.1142/S0217979224502837}} &  \multirow{2}{*}{\centering\xmark} & \multirow{2}{*}{Epidemiology} & \multirow{2}{*}{Mis/Disinformation diffusion} &  \multirow{2}{*}{4} &  \multirow{2}{*}{7.02\%}\\ 
         & & & & & \\
         \hline
         \multirow{2}{=}{\centering\cite{10.1109/GLOBECOM46510.2021.9685197,6637343,10002219}} &  \multirow{2}{*}{\centering\xmark} & \multirow{2}{*}{Game-theoretic}  & \multirow{2}{*}{Belief updating} &  \multirow{2}{*}{3} &  \multirow{2}{*}{5.26\%}\\
         & & & & & \\
         \hline
         \multirow{2}{=}{\centering\cite{10.1007/978-3-031-05637-6_35,10.1145/3603163.3609057,10.1145/3581783.3612704}} & \multirow{2}{*}{Technical} & \multirow{2}{*}{\xmark} &   \multirow{2}{*}{\xmark}   & \multirow{2}{*}{3} & \multirow{2}{*}{5.26\%}\\
         & & & & & \\
         \hline
         
         \multirow{2}{=}{\centering\cite{LIN2023105312,9381446,ButtsBollmanMurillo2023}} & \multirow{2}{*}{\centering\xmark} & \multirow{2}{*}{Information diffusion}  &  \multirow{2}{*}{Countermeasures evaluation} & \multirow{2}{*}{3} & \multirow{2}{*}{5.26\%}\\ 
         & & & & & \\
         \hline
         \multirow{2}{=}{\centering\cite{10002219,6637343}} & \multirow{2}{*}{\xmark} & \multirow{2}{*}{Game-theoretic} & \multirow{2}{*}{Mis/Disinformation diffusion} & \multirow{2}{*}{2} & \multirow{2}{*}{3.51\%}\\
         & & & & & \\
         \hline
         \multirow{2}{=}{\centering\cite{10.1007/978-3-031-35927-9_28,10.14722/ndss.2023.24293}} & \multirow{2}{*}{\xmark} & \multirow{2}{*}{Opinion dynamics}  &  \multirow{2}{*}{\xmark} &  \multirow{2}{*}{2} &  \multirow{2}{*}{3.51\%}\\
         & & & & & \\
         \hline
         \multirow{2}{=}{\centering\cite{Gavric_2019,doi:10.1177/10755470231207611}} & \multirow{2}{*}{\xmark} & \multirow{2}{*}{Epidemiology} & \multirow{2}{*}{\xmark} & \multirow{2}{*}{2} & \multirow{2}{*}{3.51\%}\\
         & & & & & \\
         \hline
         \multirow{2}{=}{\centering\cite{Terp2022,EURECOM+7114}} & \multirow{2}{*}{Technical and Social} & \multirow{2}{*}{\centering\xmark} &  \multirow{2}{*}{\centering\xmark} &  \multirow{2}{*}{2} & \multirow{2}{*}{3.51\%}\\
         & & & & & \\
    \hline
     & & & & 49 & 85.97\%\\
    \hline
    \end{tabular}
    \caption{Distribution of frameworks, models and simulation categories proposed more than once in studies.}
\end{table*}

\subsection {What purposes motivate the existing frameworks, models and simulations around mis/disinformation? (RQ3)} 

RQ3 has as its objective to establish the main purpose of the proposal of each study. Based on the review, the purposes of the frameworks, models, and simulations to understand mis/disinformation can be classified into the five following categories: mis/disinformation conceptualization, effectiveness evaluation, cognitive assessment, and mis/disinformation detection.

\begin{enumerate} 

\item {\rqthree{Mis/Disinformation conceptualization}}
Mis/Disinformation conceptualization refers to defining and structuring the fundamental concepts and elements of mis/disinformation. This involves identifying key attributes, characteristics, and mechanisms of mis/disinformation propagation and dissemination and analyzing its impact on individuals or communities. These frameworks, models, and simulations aim to increase the growing literature on the analysis and conceptualization of mis/disinformation in information environments.

For instance, the main goal of the framework proposed in \cite{Agarwal2021} was to help researchers and information professionals understand the fake news phenomenon and ways to fight it. It also helped design research studies to investigate the phenomenon of mis/disinformation empirically. Moreover, in \cite{osti_10275066}, the main purpose of their model and simulation was to help identify the impact of credible message dissemination patterns on the overall exposure to mis/disinformation on Twitter. 

Finally, in \cite{Tambuscio2001} Tambuscio \textit{et al.} conceptualized the mis/disinformation phenomenon, in their model and simulations, taking into account the combination of selective exposure due to network segregation, forgetting (i.e., finite memory), and fact-checking. 

\item {\rqthree{Effectiveness evaluation}}
In the context of mis/disinformation, effectiveness refers to the ability of proposed interventions and strategies to counteract or mitigate the impact of mis/disinformation on society. This includes assessing how these measures succeed in reducing the spread of mis/disinformation. These studies evaluate the effectiveness of mis/disinformation interventions involving countermeasures proposals or attacks. Countermeasures include a wide range of strategies, from technical solutions to algorithmic interventions to human-based approaches, offering solutions to tackle the problem of mis/disinformation targeted as increased media literacy and critical thinking skills will be addressed. However, despite the efforts of scientists and experts in the field, many challenges remain regarding mitigation, response, education, and awareness of mis/disinformation. 

For example, in \cite{Kirk2020}, the author discussed and evaluated, using simulations, different approaches, such as social constructionism, linguistics, heuristics, and empiric methods, to address the spread of mis/disinformation. Furthermore, the authors in \cite{Pourghomi2018TheSO} proposed an authentication method to reduce the spread of mis/disinformation on social media that consists of applying a verification system that can be accessed by right-clicking and providing a two-dimensional simulation for a proof-of-concept.

As mentioned above, there are not many studies that investigate mis/disinformation from an attack perspective. While a large body of research has focused on detection, mitigation, and countermeasures, studies explicitly framing mis/disinformation as a targeted attack are scarce. We identified only one study \cite{9004942} described above whose model and simulation main purpose was to evaluate the impact of different mis/disinformation attacks.

\item {\rqthree{Cognitive assessment}}
We have already mentioned the importance of considering human cognition in the context of information processing, belief formation, and susceptibility to deceptive information. These studies aim to study and analyze these aspects of mis/disinformation. 

For example, in \cite{38729e78dce24d22b6cf06e4b13384b5}, the authors analyzed the effect of large-scale confirmation bias on social learning and examined how it can drastically change the way a networked, decentralized society processes information. Moreover, the authors in \cite{10002219} analyzed how people digest information and how mis/disinformation exposure affects this process.

\item {\rqthree{Mis/Disinformation detection}}
A significant amount of work has been dedicated to the main purpose of mis/disinformation detection. Studies focusing on mis/disinformation detection aim to develop new models to distinguish between authentic and misleading or false information. By way of example, the main purpose of the EU platform CALYPSO \cite{10.1007/978-3-031-05637-6_35} is to detect and fact-check suspected mis/disinformation cases in real-time early. This platform aimed to involve the audience in promptly detecting harmful content.

Moreover, the authors in \cite{Zhou2007AnOM} aimed to demonstrate that their ontology-supported mis/disinformation model can enhance mis/disinformation detection.

Furthermore, in \cite{10.1145/3603163.3609057}, the authors proposed a framework called ``LIEALERT'', which combines advanced NLP techniques, a graph convolutional network, and attention mechanisms to detect mis/disinformation, predict its propagation depth, and cluster users based on their reactions to false information.
\end{enumerate}

Only one study was identified that was coded with two types of the above-described purposes simultaneously. The DISARM framework, described in the work by Terp \textit{et al.} \cite{Terp2022}, is a notable example of a dual-purpose approach. DISARM not only aims to increase the understanding of mis/disinformation processes but also seeks to provide practical guidance on mitigating the impact of mis/disinformation.


The main purpose of the selected paper is \rqthree{Effectiveness evaluation} (21 studies, 36.21\%). Followed by \rqthree{Mis/Disinformation conceptualization} (18 studies, 31.03\%), and \rqthree{Cognitive assessment} (10 studies, 17.24\%). Finally, \rqthree{Mis/Disinformation detection} is less common (9 studies, 15.52\%).


\subsection {Which contexts are addressed by the existing frameworks, models and simulations around mis/disinformation? (RQ4)}

Mis/disinformation can be present in very different environments. Our analysis reveals six main contexts where mis/disinformation has been present: 

\begin{enumerate}
    \item{\rqfour{Health}}
    In the context of health, mis/disinformation encompasses false or misleading information deliberately spread to influence beliefs, attitudes, decisions, or behaviors related to health and medical matters. For instance, vaccine hesitancy, mis/disinformation about COVID-19, and food safety are major concerns that directly affect the well-being of individuals and communities. In this context, mis/disinformation represents a significant threat that can directly affect public health and consumer confidence.

    For example, the authors in \cite{10.4018/IJSWIS.300827} tweets related to COVID-19 vaccination are used to perform the experiments and discuss the control strategy to minimize the misinformation and disinformation related to vaccination.

    In the same way, the model proposed in \cite{doi:10.1177/10755470231207611} adapted the Stimulus-Organism-Response (S-O-R) model where COVID-19 vaccine mis/disinformation exposure and information overload serve as environmental stimuli, perceived vaccination benefits, and barriers as cognitive organisms, fear as an affective organism, and information avoidance as a behavioral response.

    Lastly, in \cite{doi:10.1080/01292986.2020.1811737}, the authors explore the cognitive, affective, and environmental factors affecting the acquisition and diffusion of food safety mis/disinformation and show how mis/disinformation about food safety has become a serious problem in China.

    \item{\rqfour{Politics}}
    Mis/Disinformation in politics encompasses deliberately spreading false or misleading information to alter public opinion, shape political narratives, influence voting behaviors, and ultimately impact electoral outcomes. The objectives of political mis/disinformation can vary widely. They may include discrediting political opponents, manipulating public perceptions of policies or candidates, creating social divisions, fostering distrust in democratic institutions, and undermining the integrity of electoral processes.
    
    We can see an example in the model proposed in \cite{9004942}, where the authors analyzed tweets from the 2018 elections in the US when Donald Trump became the president and tweets from the Swedish mid-term elections to compare the updating of opinions.
    
    Moreover, in \cite{Zhu2023}, the authors proposed a framework for analyzing mis/disinformation policies in the United States and China, identifying various dimensions of mis/disinformation policies, including agents, objectives, use of technology, governmental actions, and social context.

    \item{\rqfour{Defense}}
    Mis/Disinformation in defense encompasses intentionally disseminating false or misleading information to undermine the integrity and effectiveness of defense operations. The objectives of using mis/disinformation in defense contexts can include deceiving adversaries about military capabilities, intentions, or strategies; creating confusion or chaos within enemy ranks; concealing actual military plans or movements; sowing distrust or misinformation among allied forces; and influencing public perception or support for military actions or policies.
    
    For example, the framework proposed in \cite{10083447} was applied to analyze mis/disinformation as a significant factor in hybrid threats, impacting the information environment and challenging distinguishing truth from lies. Hybrid threats typically combine conventional and unconventional methods, blending military and non-military tactics to achieve strategic objectives.

    Moreover, in \cite{Kryshtanovych2023AnIM}, the authors proposed a framework for counteracting mis/disinformation's negative impact on the cybersecurity system. This framework aims to combat mis/disinformation, strengthen the defense mechanisms of cybersecurity systems, and protect against potential threats posed by mis/disinformation and cyber attacks.

    Lastly, the framework mentioned above proposed in \cite{Terp2022} called ``DISARM'' aims to manage information security-based standards for detecting and responding to information harms, including mis/disinformation.
    
    \item{\rqfour{News articles}}
    Mis/Disinformation in news articles encompasses the deliberate spread of false or misleading information with the intent to undermine trust in media sources, distort the public's perception of events, and contribute to the dissemination of biased narratives. Its primary objectives include influencing public opinion, shaping political discourse, and manipulating the narrative surrounding specific topics or events.
    
    For example, the framework proposed in the EU project CALYPSO \cite{10.1007/978-3-031-05637-6_35} is developing a digital platform to source citizens’ contributions to false and misleading news articles. Moreover, in \cite{Agarwal2021}, the authors review the literature on mis/disinformation, primarily from a news article perspective.
    
    \item{\rqfour{Crises}}
    In humanitarian crises, characterized by situations where communities or large populations are exposed to life-threatening conditions or imminent danger, mis/disinformation can exacerbate the severity of the crisis and lead to increased harm. Its primary objectives include spreading confusion, undermining trust in official sources of information, and impeding coordinated responses to the crisis, thereby prolonging and intensifying its impact on affected populations.
    
    Examples of such crises include global warming, which emerges as a pressing and multifaceted challenge that extends beyond environmental concerns. Mis/Disinformation can contribute to climate change since it can cause large segments of our societies to not believe in the reality of climate change.
    
    Natural disasters are another type of crisis. In this context, mis/disinformation can exacerbate the devastating consequences of earthquakes, hurricanes, floods, and other natural disasters. It can significantly impact decision-making, emergency response, and public safety.
    
    For instance, in \cite{Tran2020MisinformationIC}, the authors investigate human-machine interactions generating and mitigating mis/disinformation during humanitarian crises and propose a conceptual framework based on two activity systems: generating and mitigating mis/disinformation.
    
    Moreover, the authors in \cite{38729e78dce24d22b6cf06e4b13384b5} investigate the model's predictions empirically using US county-level data on the impact of internet access on the formation of beliefs about global warming.
    
    Finally, the authors in \cite{JIANG2021125572}, acknowledging that effective debunking strategy is a potential tool to reduce the loss of massive digital mis/disinformation, proposed a novel rumor spreading–debunking (RSD) model by ordinary differential equation (ODE) system to explore the interplay mechanism between rumor spreading and debunking processes. Using a real-world rumor case, the ``immigration rumor'' during Hurricane Harvey in 2017, for their model's simulations.
    
    \item{\rqfour{E-commerce}}
    Mis/Disinformation in the context of e-commerce encompasses the deliberate dissemination of false or misleading information related to products, services, or reviews on online platforms. This misleading information can include fake product reviews, manipulated ratings, deceptive advertising, or fraudulent product features or benefits claims. The primary objective of mis/disinformation in e-commerce is to mislead consumers, influence their purchasing decisions, and create a false perception of products or services offered on these platforms. This can lead to a lack of trust among users, reduced credibility of online reviews, decreased customer satisfaction, and financial losses for consumers and businesses.
    
    In this context, the authors in \cite{Govindankutty2023} analyzed the social and emotional intelligence of communities and consumers frequently exposed to mis/disinformation and sharing fake news and reviews.

\end{enumerate}

\begin{figure*}[ht]
    \centering
    \includegraphics[width=0.9\textwidth]{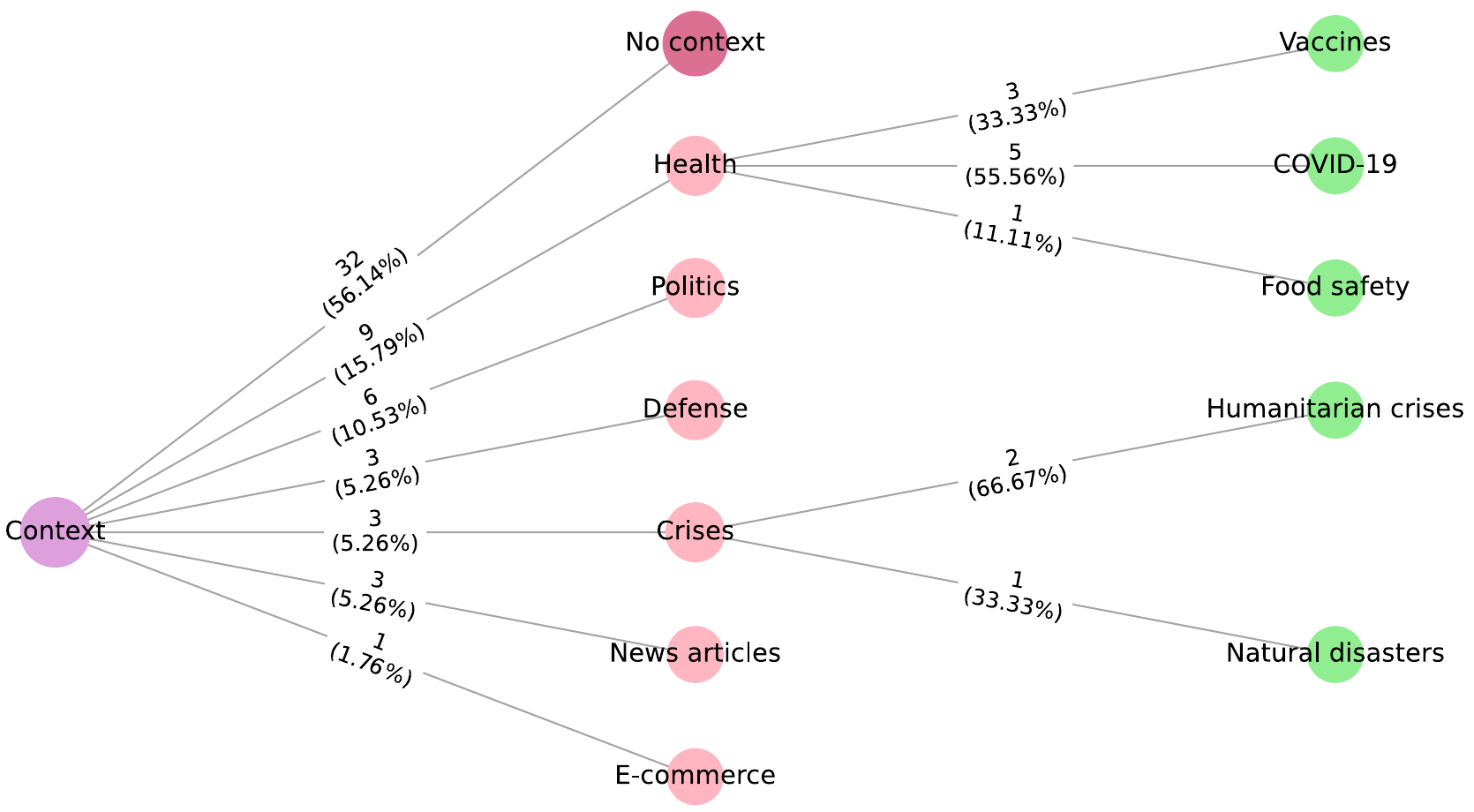}
    \caption{Contexts of analyzed mis/disinformation frameworks, models and simulations.}
    \label{fig:rq4}
\end{figure*}

Despite the different contexts identified in the analysis, most studies did not specify in what context mis/disinformation was used (32 studies, 56.14\%). Figure \ref{fig:rq4} presents the distribution of the environments across the selected articles. \rqfour{Health} and \rqfour{Politics} are the most employed context (9 studies, 15.79\%). Followed by \rqfour{Defense}, \rqfour{Crises} and \rqfour{News articles} (3 studies, 5.26\%). \rqfour{E-commerce} is the less employed context (1 study, 1.76\%)

\subsection {Which validation is performed in the existing frameworks, models and simulations around mis/disinformation? (RQ5)}

Validation refers to confirming the accuracy, reliability, and applicability of a framework, model, or simulation by comparing and thoroughly analyzing it against independent external data. Validation is relevant for establishing the reliability and robustness of methods, models, and frameworks designed to address the challenge of mis/disinformation. From our analysis, we find five primary categories:

\begin{enumerate}
    \item\rqfive{Simulations with real-world data}
    Simulations with real-world data refer to a validation technique where frameworks and models are tested using actual data collected from the real world. This process involves running simulations based on authentic datasets to assess how well the framework or model performs in replicating real-world scenarios and outcomes.
    
    For example, in \cite{10.1109/GLOBECOM46510.2021.9685197}, the authors take a sample of 1,000 users from a real Twitter dataset and evaluate the proposed game-theoretic opinion model, using these data to simulate five different types of networks (uncertainty, homophily, assertion, herding, and encounter-based) to analyze how each opinion model handles mis/disinformation.
    
    Moreover, the authors in \cite{ABOUZEID2022100341} evaluate the proposed framework and models' impact on social network data like Twitter and Facebook, demonstrating effectiveness in simulating information diffusion and identifying mis/disinformation mitigation strategies.
    
    Finally, in \cite{6637343}, the authors used extensive simulation based on a real-world online network data set (with 1,899 nodes) and another real-world dataset (with 9,877 nodes). To validate their game-theoretic model in the ﬁrst effort to use the concepts of `Evolutionary graph theory' in game-theoretical applications to study the information diffusion process in social networks. To analyze and predict the spread of mis/disinformation.
    
    \item{\rqfive{Comparing with real-world data}}
    Comparing real-world data involves verifying the outcomes or predictions generated by frameworks or models with empirical observations obtained from real-world sources. This validation technique assesses the consistency between simulated results and actual data to determine the framework or model's accuracy and reliability. 
    
    For instance, in \cite{Liu2022}, the authors proposed a model that addresses mis/disinformation in emergency events by considering it a key factor in the formation of multiple opinions in online social networks to analyze how mis/disinformation affects online opinion dynamics during emergency events and how different official responses can intervene to manage evolving public opinion. This proposed model is validated by comparing simulation results with real-world data collected from online behaviors on the ``Sina Weibo'' platform during emergencies. Data on user interactions, such as likes on comments, reposts, and user profiles, are collected.
    
    \item{\rqfive{Surveys}}
    Surveys are quantitative research methods that consist of collecting data from people to evaluate methods' effectiveness and obtain valuable information on the acceptability and usefulness of theoretical and practical proposals. Typically, surveys consist of questions designed to elicit responses from a specific sample of participants.
    
    For example, in \cite{9920336}, the authors employed the DISCERN questionnaire to evaluate their framework for analyzing YouTube content recommendation paths to monitor and detect health mis/disinformation on YouTube by assessing content networks and identifying key nodes within those networks.
    
    In the same way in \cite{LIN2023105312}, Lin \textit{et al.} used surveys to measure the effectiveness of proposed countermeasures, simulating the drift-diffusion model (DDM) to investigate the impact of accuracy prompts on individuals' sharing intentions. The score is based on the credibility of the news.
    
    Moreover, the authors in \cite{10.14722/ndss.2023.24293} explored different `folk' models of mis/disinformation on social media to understand individuals' perceptions and responses to mis/disinformation. The model has been validated through surveys, which consist of questions that addressed issues such as the origin of mis/disinformation, the perceived purpose of mis/disinformation, users' responses to mis/disinformation, and how they evaluate whether or not mis/disinformation.
    
    \item{\rqfive{Focus groups}}
    Focus groups are a type of research method in which a small group of people discuss a specific topic guided by a trained moderator. This group interaction helps researchers uncover deeper reasons for people's opinions and behaviors. It is a valuable tool for exploring why people think or feel a certain way.
    
    For example, in \cite{Tran2020MisinformationIC}, Tran \textit{et al.} proposed a framework that focuses on examining human-machine interactions in the context of mis/disinformation generation and mitigation and validated their framework by engaging two groups of graduate students with two defined scenarios of humanitarian crises' mis/disinformation to validate the conceptual framework. 
    
    \item{\rqfive{Mathematical proof}}
    To measure the correctness of a mathematical model, it is not necessary to use real data. Theoretical model proposals can be mathematically proven. By way of example, the authors in \cite{Govindankutty2023} propose a model called ``SEDIS'', which includes states such as Susceptible, Exposed, Doubtful, and Infected, reflecting how people react to mis/disinformation online, and evaluate the proposed model by solving different equations by varying values of the parameters that form it.
    
    Furthermore, in \cite{10.1007/978-3-031-35927-9_28} the authors proposed an opinion dynamics model to examine the relationship between users' personality traits such as extroversion, agreeableness, conscientiousness, and neuroticism and their engagement with mis/disinformation on social media during the COVID-19 pandemic. The proposed model has been validated through multinomial logistic regression analysis to predict mis/disinformation engagement from social characteristics and personality traits. 
    
    Moreover, in \cite{GhoshDas2023}, the authors extended the traditional Susceptible-Exposed-Infected-Recovered (SEIR) model to study the dynamics of propagation of online mis/disinformation, considering four categories of users of social networking sites, namely, ignorant population, believers, active spreaders and stiflers. The model has been validated by determining the critical value of the spread of mis/disinformation, which regulates the stability switch from a state without spreaders to a state with spreaders.

\end{enumerate}

Having analyzed the articles of this survey, nearly half of them (25 studies, 43.86\%) did not validate their proposal. On the contrary, validations through \rqfive{Simulations with real-world data}, and \rqfive{Comparing with real-world data} are the most common (11 studies, 19.30\%). Followed by \rqfive{Surveys} (6 studies, 10.53\%). The less common ways of validation are \rqfive{Mathematical proof} (3 studies, 5.26\%) and \rqfive{Focus groups} (1 study, 1.75\%).



To analyze the relationship between the type of validation (Simulations with real-world data, Comparing with real-world data, Surveys, Mathematical proof, Focus group) and the type of validated element (Framework, Model, Simulation), we employ a chi-square test of independence to determine if there is an association between them.

The analysis involved creating a contingency table and performing a chi-square test to determine if there was a significant association between the variables. However, the results revealed a high $p-value$ of 0.816, suggesting insufficient evidence to reject the null hypothesis of independence. Therefore, we cannot conclude that there is a significant relationship between the type of validation and the type of validated element in this dataset. These findings imply that the type of validation does not exhibit a meaningful association with the type of validated element, highlighting potential complexities or nuances in the validation processes across different contexts or domains.

Regarding combinations that appeared more than twice, excluding not validated approaches. The most frequent combination (4 appearances) is validated epidemiology models and belief updating simulations with simulations with real-world data. The last combination appears twice and is validated again by epidemiology models and belief updating simulations in this case by comparing with real-world data.

\section{Findings analysis} \label{discussion}

\begin{figure*}[ht!]
    \centering
    \includegraphics[width=\linewidth]{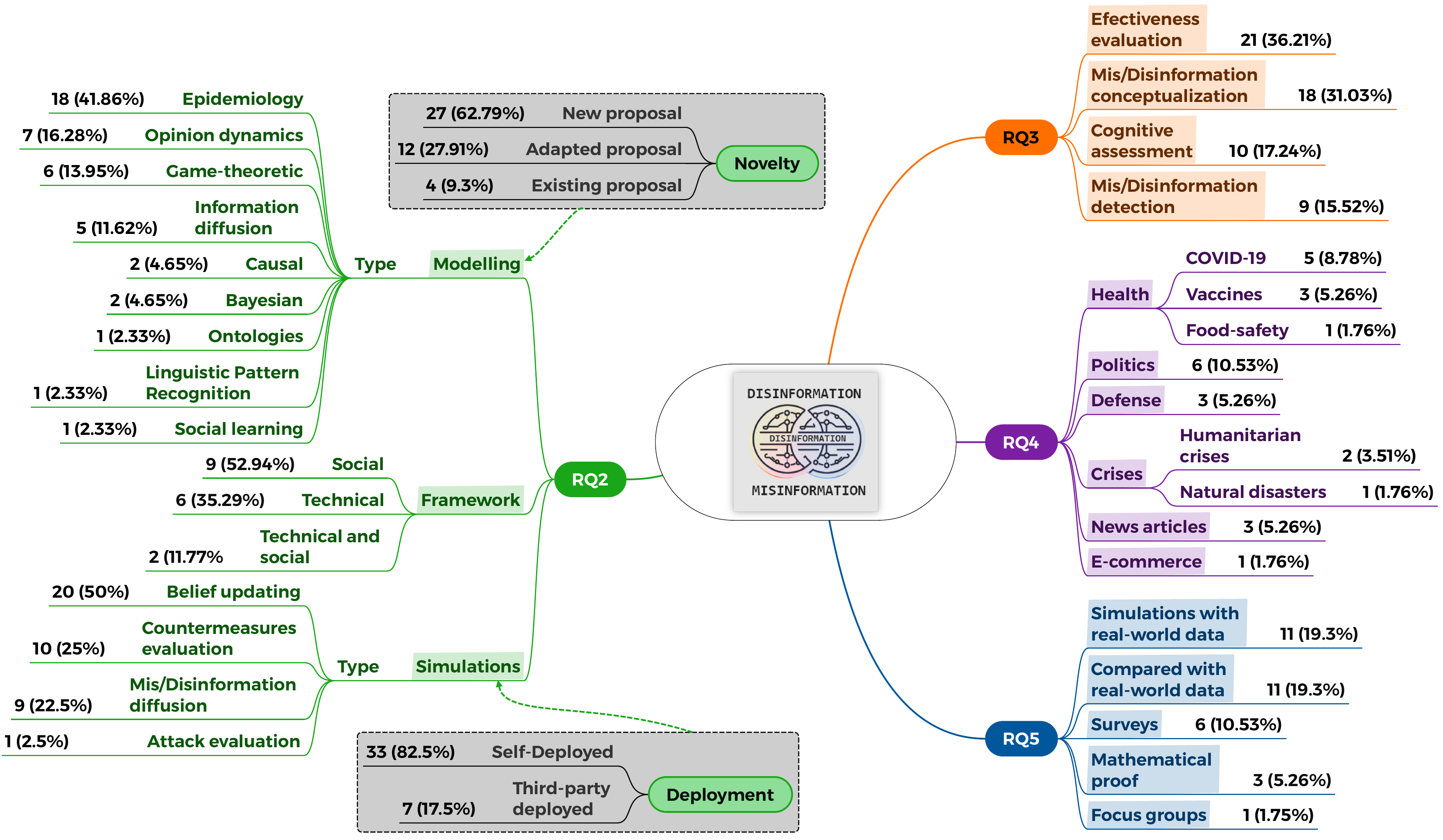}
    \caption{Summary of the main findings from the literature of frameworks, models and simulation along the mis/disinformation phenomena.}
    \label{fig:summary}
\end{figure*}

In this section, we provide a summary of our key findings, with a visual representation encapsulated in Figure \ref{fig:summary}, which shows the distribution of papers in each category along the RQs explored in the paper. Next, the past and future challenges in frameworks, models and simulation along the mis/disinformation phenomena are discussed. Finally, the existing limitations in this study and the implications of our research are addressed.

\subsection{Current trends}

The current trends in frameworks, models, and simulations within the mis/disinformation phenomenon extracted from the 57 papers analyzed are described below to answer each RQ proposed in this study.

\subsubsection{Mis/Disinformation terminology (RQ1)}

First, we analyzed the definitions and concepts the studies applied to mis/disinformation (RQ1), finding that most provided a definition, which can be: misinformation refers to unintentionally false or inaccurate information, while disinformation entails deliberately spreading false or misleading information with the intent to deceive or manipulate. Still, many studies did not provide a definition or a definition unrelated to intentionality, generating controversy around the concept of disinformation. In this sense, Kapantai \textit{et al.} \cite{Kapantai2020ASL} emphasizes the importance of clear and commonly accepted definitions since different disinformation types might require different theoretical analyses and one of their objectives was to identify and organize the diverse definitions of the concept. In this study, the term ``disinformation'' has been used to refer to information that meets the European Commission's definition. This term is preferred to ``fake news'' or ``misinformation'' because it is more accurate and covers a wider range of misleading, inaccurate, or false information.

\subsubsection{Mis/Disinformation representations: Frameworks, models, and simulations (RQ2)}

Across papers, researchers used many different representations (RQ2) to mis/disinformation. We classified them into three main categories that are frameworks, models and simulations. We found that \rqtwo{epidemiology} is the most common model. While epidemiology is particularly suitable for capturing the dynamics of mis/disinformation since it can represent the spread of mis/disinformation and the change in people's beliefs \cite{Kryshtanovych2023AnIM}. It is important to accept its assumptions and inherent limitations, such as the notion of information `contagion' and the possibility of oversimplification of complex social and political dynamics. One of the most significant discoveries we have is the relationship between models, simulations, and frameworks. We found that model proposals were combined with simulations in order to prove the proposed model. On the other hand, frameworks usually appear without models or simulations. These findings align with the expectation that an association with the use of a framework is typically linked to a conceptual framework without proposed models or simulations.

Unlike models, which predominantly focus on representing the dynamics of mis/disinformation spread, frameworks serve as organizational structures encompassing various aspects of mis/disinformation. We found that only a small portion of the studies propose a framework and most of them proposed social and high-level approaches.

In this scenario, few frameworks address all aspects of mis/disinformation, from detection to mitigation. One of the most complete is DISARM \cite{Terp2022}; the EU, NATO, and the ONU are organizations that support it. Others, like the proposed by Kapantai \textit{et al.} \cite{Kapantai2020ASL}, emphasize the different mis/disinformation types. The authors proposed three independent dimensions with controlled values per dimension as categorization criteria for all types of mis/disinformation but did not propose any countermeasure. This absence of countermeasures also occurs in the SCOTCH framework \cite{atlanticcouncilSCOTCHFramework}. This framework was developed closely with the United States Government (USG). Unlike the two previous ones, it proposes an analysis of the actors involved in the dissemination of mis/disinformation. This suggests a potential gap in the literature concerning the development of frameworks that holistically encompass the diverse dimensions of the mis/disinformation phenomenon.

Concerning simulations, we found that \rqtwo{belief updating} is the most common simulation type for mis/disinformation, followed by \rqtwo{countermeasures evaluation}. There is a strong interest in investigating mis/disinformation's effects on individual and collective beliefs, as evidenced by the widespread use of \rqtwo{belief updating} simulations.

Moreover, the focus on \rqtwo{countermeasures evaluation} simulations indicates a concerted effort to assess strategies to mitigate mis/disinformation's impact. We also noticed that few studies mainly aimed to simulate mis/disinformation attacks (\rqtwo{Offensive evaluation}), specifically only one study. This reflects a greater emphasis on developing defense and mitigation strategies, prioritizing identifying and countering existing mis/disinformation rather than actively simulating new attacks.

\subsubsection{Mis/Disinformation purposes (RQ3)}

Regarding the purposes (RQ3), \rqthree{efectiveness evaluation} is the most common purpose. This observation highlights the research community's efforts in developing effective strategies and interventions to evaluate and combat the negative impact of mis/disinformation. In addition to the scientific community's dedication to addressing mis/disinformation with countermeasures, it is also important to recognize the involvement of external entities and organizations in this crucial effort. A prime example is the EU's active engagement in this fight against mis/disinformation, as seen in ``EU vs Disinfo'' proposed in 2015 \cite{euvsdisinfoINICIODisinfo}.

\subsubsection{Mis/Disinformation contexts (RQ4)}

We also extracted six main contexts (RQ4) across studies. We found that most of them occurred in \rqfour{health}, especially in \rqfour{COVID-19}, which shows that this context is a fertile ground for possible mis/disinformation campaigns. With the same number of studies, we have \rqfour{politics}, underscoring the pervasive nature of political mis/disinformation in shaping public discourse and opinion formation, with potentially far-reaching implications for democratic processes and societal cohesion.

Furthermore, most studies did not specify the context where mis/disinformation was applied. However, due to the limitations imposed by our inclusion and exclusion criteria, mis/disinformation will always be approached from a digital context. We have focused only on academic studies and current contexts where mis/disinformation is critical, such as wars, political conflicts, etc. This can lead to a biased view of the problem since mis/disinformation is also present in everyday contexts, such as sports, where its impact is not as critical.

\subsubsection{Mis/Disinformation validation methods (RQ5)}

Finally, we wish to report the extent of validation of the approaches (RQ5). We found that half of the studies had not validated their proposal, and most of the validated studies performed \rqfive{simulations with real-world data} or compared their results with these types of data (\rqfive{comparing with real-world data}). This suggests a trend utilizing realistic scenarios or benchmarks for validation, highlighting the importance of linking theoretical proposals with real-world situations.

\subsection{Open challenges}

Based on our findings and previous related studies, we find some open challenges in the area authors usually report. A description of each of these challenges is found below:

\subsubsection{Standardizing mis/disinformation frameworks, models and simulations}

An important open challenge is establishing a standardized and universally accepted mis/disinformation model to facilitate study comparability and interoperability. The same applies to frameworks. Our review found that DISARM \cite{Terp2022} was the most complete approach with the European Union's support. Still, their authors mentioned that the framework should be tested in more scenarios and expanded with more incident types. Moreover, more proposals aim to cover new areas, like the analysis of the actors involved in the dissemination of mis/disinformation \cite{atlanticcouncilSCOTCHFramework}. Future research should focus on developing new frameworks that include a mis/disinformation model. These new frameworks could be the combinations or extensions of available frameworks that should be tested and validated.

One possible way to test frameworks and models can be real scenario simulations. Based on our findings, we noted that, at present, most studies that proposed mis/disinformation frameworks or models applied simulations of their approach. However, we identified a few papers that fully utilize newer and more powerful technologies, such as GenAI, particularly LLMs, which do not need real-world data to simulate mis/disinformation scenarios. For example, the authors in \cite{pastorgalindo2023generative} discussed the potential of LLMs as an innovative agent-based method for comprehending, simulating, and evaluating mis/disinformation within controlled experimental settings. Thus, future research should consider LLMs' affordances in mis/disinformation research.

\subsubsection{Contextual variability in mis/disinformation frameworks, models and simulations}

Mis/Disinformation can be spread in any context and situation. For example, Capuano \textit{et al.} \cite{Capuano2023} indicated that most mis/disinformation datasets are about politics and highlighted the need for a standard method for storing mis/disinformation data with more diverse topics. Moreover, the importance of the context is highlighted in \cite{slr1}. The authors focused on fake news detection and discussed the challenge of how models with high accuracy on one concrete topic do not work as well as on another topic. Future mis/disinformation proposals should be independent of context to adapt to a dynamic and changing environment like digital media. 

\subsubsection{Education as countermeasure}

Regarding the purposes, we found that `countermeasures' are the most common purpose in our results. However, we identified the challenges in implementing tools that aim to eradicate mis/disinformation at its root by focusing on educating and training individuals. Among our findings, only the CALYPSO platform \cite{10.1007/978-3-031-05637-6_35}, part of an EU project, attempts to involve the public in mis/disinformation detection. However, it is currently in a prototype stage, with its sole function being to have individuals classify content as true, false, or undefined.

A potential solution to educate the population about mis/disinformation could involve incorporating serious games with this theme. Although we did not find any proposals among our results, there are indeed serious games addressing mis/disinformation available. For instance, Tilt was established to enhance resilience against online manipulation. One of their offerings is the serious game called ``Harmony Square'' \cite{tiltstudioGameCombat}, centered around the theme of fake news. The game unfolds in the idyllic Harmony Square, a small neighborhood with a mild obsession with democracy. As the player, you assume the `Chief Disinformation Officer' role. Throughout four brief levels, your task is to disrupt the square's peace by fostering internal divisions and turning its residents against each other. The game aims to expose tactics and manipulation techniques used to mislead people, garner a following, or exploit societal tensions for political purposes.

Another potential solution involves using Cyber Ranges, a platform that simulates real operational environments for the individual or collective training of professionals. While no proposals were identified in the literature, the FFI (Norwegian Defence Research Establishment) is currently developing ``Somulator'' \cite{ffiSomulator}, a solution designed to simulate social networks in exercises, capable of simulating online newspapers, microblogs, image sharing, video sharing, and more generic social networks.

Moreover, in 2024, the NATO Strategic Communications Centre of Excellence (NATO StratCom COE) has unveiled its latest initiative to revolutionize military and strategic communications (StratCom) training. Known as the Information Environment Simulation Range (InfoRange) \cite{natostratcomcoe}, this platform offers an immersive simulated training environment that leverages advanced technology to elevate tabletop exercises and provide comprehensive crisis or conflict training in a responsive digital information environment. The InfoRange incorporates generative AI to facilitate exercise design and audience application, offering dynamic training opportunities by simulating realistic information environment infrastructure, with the aim to training audiences across various StratCom sub-disciplines, including public diplomacy, civilian and military public affairs, InfoOps, and PsyOps.

\subsection{Survey limitations and implications}

This study has inherent limitations, which may impact our findings:

\subsubsection{Paper selection}

The paper selection mainly limits this review. First, we have only used the key terms ``disinformation'' and ``misinformation'' to perform our document search based on the papers' titles. Other studies could also be working on disinformation, but they might use slightly different terms to describe their work. As we have seen in RQ1, there is no consensus on the terminology of the phenomenon. Therefore, those studies might not be included in our review. Nevertheless, we purposely opted for these terms to analyze the core of disinformation while having a manageable selection of papers for this study. Furthermore, we focused on the primary academic databases of Scopus and Web of Science. However, other peer-reviewed academic papers could be indexed in different databases and non-peer-reviewed publications, including pre-prints, technical or white reports that could be missing in our review, and non-academic work being conducted in industrial companies and by practitioners. Finally, we have based our RQ generation with a focus on frameworks, models, or simulations related to mis/disinformation. Still, other potential and valuable RQs about the mis/disinformation phenomenon might be missing in this review.

\subsubsection{Ethical concerns}

The study of mis/disinformation raises ethical challenges regarding the social responsibility of the media and freedom of expression \cite{mathiesen2019fake}. Regulating mis/disinformation also poses challenges, as it can lead to sensitive situations, such as censorship of information and non-censorship of mis/disinformation. Delving into the regulation of mis/disinformation introduces its own set of complexities. A balance is needed, as strict rules can inadvertently censor legitimate information, undermining the principle of free expression, while inappropriate measures may allow mis/disinformation to proliferate unchecked. Another challenge is navigating the subjective nature of distinguishing truth from falsehood \cite{139062f0-c933-3a5e-86c2-25e4c2cbcc4d}. Recognizing that determinate truth often requires the establishment of a subject, attempts to establish universal truths raise concerns about the imposition of a single narrative or perspective. A balance must be struck against avoiding further imposition of a single truth.

\section{Conclusion} \label{conclusion}

This study presents a systematic literature review encompassing frameworks, models, and simulations proposed up to 2023 for investigating the phenomena of misinformation and mis/disinformation. To our knowledge, this paper stands as the pioneering effort in reviewing methodologies aimed at representing and comprehending the dynamics of misinformation and mis/disinformation. Utilizing the widely recognized PRISMA methodology, we comprehensively searched the two foremost bibliographic databases, Scopus and Web of Science. Following stringent selection criteria, 57 papers were identified for further in-depth analysis to respond to five research questions (RQ).

Firstly, RQ1 focused on exploring the terminology and definitions utilized to describe misinformation and disinformation, as the absence of an official definition necessitates clarification in the academic discourse. We concluded that there is a common consensus regarding the definitions of disinformation and misinformation, which revolves around intentionally disseminating false information. Particularly, disinformation refers to false information deliberately spread to manipulate or influence, while misinformation refers to false information without intent to deceive.

Secondly, RQ2 delves into the formal representation of mis/disinformation within frameworks, models, and simulations, shedding light on how this complex phenomenon is conceptualized and operationalized. Among the identified frameworks, most studies propose social frameworks, while a subset focuses on technical frameworks; intriguingly, only two studies integrate both approaches. Epidemiology models emerge as the predominant approach in modeling mis/disinformation, leveraging them to simulate propagation akin to the spread of disease within populations, often introducing novel models that incorporate additional or diverse agent states. Opinion dynamics models follow closely, offering innovative approaches to representing mis/disinformation and analyzing the evolution of opinions surrounding it. Game-theoretic models and information diffusion models are also present in the landscape. In terms of simulations, belief updating stands out as the predominant simulation type employed in mis/disinformation studies, followed by countermeasures evaluation. Mis/Disinformation diffusion is also utilized alongside offensive evaluation, which appears in only one study.

RQ3 delves into the primary objectives of the research being done through the frameworks, models, and simulations. The proposals' purposes are related to evaluating the effectiveness of mis/disinformation attacks and associated countermeasures, formalizing mis/disinformation, assessing cognitive impact, and detecting mis/disinformation. 

RQ4 aims to identify the contexts where misinformation and mis/disinformation are most prevalent, offering valuable insights into the domains and subjects of attention from the research community. Some studies and solutions are mainly proposed in the context of health and politics, but also defence, crises, news articles and e-commerce are areas for mis/disinformation research. However, most proposals were generic and agnostic from the field of application, with no context being defined.

Finally, RQ5 examines the methods employed to validate the proposed frameworks, models, and simulations to ensure the feasibility and reliability of the approaches. A significant portion did not validate their proposals was observed. Nevertheless, among the studies that did validate their frameworks, models and simulations, most studies validated their proposals using simulations or comparisons with real-world data, followed by surveys. Conversely, mathematical proof and focus groups were less commonly used for validation.

Mis/Disinformation has become an integral aspect of our digital lives. It has already proven to be potentially very dangerous in the digital ecosystem and outside of it, especially with the rise of fake news, memes, deepfakes, and emotionalism. 

In such a scenario, a trend in the model and simulating mis/disinformation offers insights into its dynamics and evolution and evaluates the effectiveness of mis/disinformation attacks and countermeasures. However, a holistic approach with models, frameworks, and simulations that consider social, psychological, technological, and contextual factors is lacking. Additionally, these approaches should be validated rigorously, ensuring their reliability and accuracy. Finally, these holistic approaches should prioritize empowering individuals at the heart of the issue by providing them with the necessary training and resources to detect and combat mis/disinformation effectively. 

Future work in the mis/disinformation area may prioritize standardizing frameworks, models and simulations to enhance comparability and interoperability across studies, adapting to diverse contexts and situations. Additionally, educational interventions to empower individuals to identify and mitigate the impact of mis/disinformation, leveraging innovative approaches such as serious games, need to be developed.


\begin{acks}

This work has been partially funded by the strategic project CDL-TALENTUM from the Spanish National Institute of Cybersecurity (INCIBE), the Recovery, Transformation, and Resilience Plan, Next Generation EU, and the University of Murcia by FPU contract.

\end{acks}

\bibliographystyle{ACM-Reference-Format}
\bibliography{sample-base}

\end{document}